\documentclass[%
 aip,
 jcp,%
 amsmath,amssymb,
 reprint,%
]{revtex4-1}

\usepackage{graphicx}
\usepackage{dcolumn}
\usepackage{bm}
\usepackage{dsfont}
\usepackage{placeins}
\begin{document}
\preprint{AIP/123-QED}

\title[Analysis of the Adaptive Multilevel Splitting method on the isomerization of alanine dipeptide]
{Analysis of the Adaptive Multilevel Splitting method on the isomerization of alanine dipeptide}
\author{Laura J. S. Lopes}\email{laura.silva-lopes@enpc.fr}
\author{Tony Leli\`evre}\email{tony.lelievre@enpc.fr}
\affiliation{CERMICS, \'Ecole des Ponts ParisTech, 6-8 avenue Blaise Pascal, 77455 Marne-la-Vall\'ee, France}
\date{\today}

\begin{abstract}
We apply the Adaptive Multilevel Splitting method to the $C_{eq} \rightarrow C_{ax}$ transition of alanine dipeptide in vacuum. 
Some properties of the algorithm are numerically illustrated, such as the unbiasedness of the probability estimator and the robustness of the method with respect to the choice of the reaction coordinate.
We also calculate the transition time obtained via the probability estimator, using an appropriate ensemble of initial conditions.
Finally, we show how the Adaptive Multilevel Splitting method can be used to compute an approximation of the committor function.
\end{abstract}

\maketitle

\section{Introduction}

Simulation of rare events has been an important field of research in biophysics for nearly two and a half decades now. 
The goal is to obtain kinetic information for processes like protein (un)folding or ligand-protein (un)binding.
A usual quantity of interest is the transition rate, or equivalently its inverse, the transition time. 
This quantity is, for example, directly related to drug-target affinity, making its calculation an important step in drug design\cite{copeland-pompliano-meek-2006}.
The committor function, which gives the probability to reach a targeted configuration before going back to the initial conformation, is also interesting for computational and modeling purposes\cite{transition-path-theory}.

The events of interest in molecular dynamics generally involve transition between metastable states, which are regions of the phase space where the system tends to stay trapped.
These transitions are rare, making the simulation too long and sometimes even computationally impracticable. 
To deal with this difficulty, sampling methods have been developed to efficiently simulate rare events.
Among them are splitting methods, that consists in dividing the rare event of interest into successive nested more likely events.
For example, a reactive trajectory is divided into pieces which gradually progress from the initial state to the target one.
Examples of splitting methods include Milestoning\cite{faradjian-elber-04}, Weighted Ensemble\cite{WE-1998}, Forward Flux Sampling\cite{FFS-2009} and Transition Interface Sampling\cite{TIS-2005}.
In these methods, the intermediate milestones or dividing surfaces, used to split the rare event of interest, are fixed, so they are parameters that should be defined in advance.
Let us however mention that there exists an adaptive version of the Forward Flux Sampling method\cite{FFS-2009}, in which a few preliminary runs enable to optimize the position of the dividing surfaces.

The Adaptive Multilevel Splitting (AMS) method\cite{cerou-guyader-07a} is a splitting method in which the positions of the intermediate interfaces, used to split reactive trajectories, are adapted on the fly, so they are not parameters of the algorithm.
The surfaces are defined such that the probability of transition between them is constant, which are known to be the best surfaces in terms of the variance of the rare event probability estimator\cite{cerou-delyon-guyader-rousset-2018}.
Moreover, as illustrated below, the method gives reliable results for a large class of sensible reaction coordinates, making it particularly straightforward to use for practitioners.
This method has been used with success to estimate rare events probabilities in many contexts.
In particular, the AMS method was already efficiently applied to a large scale system to calculate unbinding time\cite{AMS-benzamidine}.
Let us emphasize that the AMS algorithm can be used not only to estimate the probability of a rare event, but also to simulate the associated rare events (typically, the ensemble of reactive trajectories in the context of molecular dynamics).
This allows us to study the possible transition mechanisms, that are often more than one, and to estimate the committor function, for example.

Compared to previous publications on AMS\cite{AMS-benzamidine,temps-trans}, we provide in this paper a full description of the correct way to implement the algorithm in a discrete in time setting.
The reader will find this description in Section \ref{section:methods}, as well as a brief discussion of some important properties of the method and the way to obtain the transition time using AMS.
We apply the method to a toy problem, namely the isomerization of alanine dipeptide in vacuum ($C_{eq} \rightarrow C_{ax}$ transition).
In this small example, we are able to numerically illustrate the consistency and the unbiasedness of the AMS method, as well as to explore in details its properties, by comparing the results to brute force direct numerical simulation.
These numerical results are reported in Section \ref{section:numerical}.
They illustrate the interest of the method and lead us to draw useful practical recommendations to get reliable results with AMS.

\section{Methods}
\label{section:methods}

Assume that the simulations are done using Langevin dynamics.  
Let us denote by $\mathbf{X}_t=(\mathbf{q}_t,\mathbf{p}_t) \in \mathbb{R}^{d\times d}$ the positions and momenta of all the particles in the system at discrete time $t$, $d$ being three times the number of atoms.
The vector $\mathbf{X}_t$ evolves according to a time discretization of the Langevin dynamics such as:
\begin{equation}
\left\{
\begin{array}{lcl}
\mathbf{p}_{t+\frac{1}{2}} & = & \mathbf{p}_t - \dfrac{\Delta t}{2}\nabla V(\mathbf{q}_t)-\dfrac{\Delta t}{2}\gamma M^{-1} \mathbf{p}_{t}\vspace{0.1cm}\\
&& +\sqrt{\Delta t \gamma \beta^{-1}}\mathbf{G}^t \vspace{0.1cm}\\
\mathbf{q}_{t+1} & = & \mathbf{q}_t + \Delta t M^{-1} \mathbf{p}_{t+\frac{1}{2}}\vspace{0.1cm}\\
\mathbf{p}_{t+1} & = & \mathbf{p}_{t+\frac{1}{2}} - \dfrac{\Delta t}{2}\nabla V(\mathbf{q}_{t+1})\vspace{0.1cm}\\
&& - \dfrac{\Delta t}{2} \gamma M^{-1} \mathbf{p}_{t+1} + \sqrt{\Delta t \gamma \beta^{-1}}\mathbf{G}^{t+\frac{1}{2}}.
\end{array}
\right.
\label{langevin}
\end{equation}
Here, $V$ denotes the potential function, $M$ is the mass tensor, $\gamma$ is the friction parameter, $\beta^{-1}=k_B T$ is proportional to the temperature, and $(\mathbf{G}^t,\mathbf{G}^{t+\frac{1}{2}})_{t \ge 0}$ is a sequence of independent centered Gaussian vectors with covariance identity.
Let us emphasize that, although we use this dynamic as an example to present the algorithm, it applies to any Markovian stochastic dynamics (like overdamped Langevin, Andersen thermostat, kinetic Monte Carlo, etc...).

Let us call $A$ and $B$ the source and target regions of interest.
The goal is to sample reaction trajectories that link $A$ and $B$ and to estimate associated quantities.
Both~$A$~and~$B$~are subsets of $\mathbb{R}^{d\times d}$.
In practice, they are typically defined only in terms of positions.
In addition, assume that $A$ is a metastable region for the dynamics.
This means that starting from a point in the neighborhood of~$A$, the trajectory is most likely to enter~$A$~before visiting~$B$.
To measure the progress from~$A$~to~$B$~one needs to introduce a reaction coordinate $\xi$, i.e. a real-valued function defined over $\mathbb{R}^{d\times d}$, whose values will be called levels.
Again, in practice,~$\xi$~typically only depends on the positions of the atoms.
The function $\xi$ is assumed to satisfy the following condition:
\begin{equation}
\exists \text{ } z_{max} \in \mathbb{R} \text{ such that } B \subset \xi^{-1}(]z_{max},+ \infty[),
\label{cond-xi}
\end{equation}
that makes necessary to exceed a level $z_{max}$ of $\xi$ to enter~$B$ when starting from $A$.
Let us emphasize that this is the only condition we assume on $\xi$ in the following: the algorithm can thus be applied with many different reaction coordinates.

Note that the definitions of the zones $A$ and $B$ are independent of the reaction coordinate. 
Since $\xi$ does not need to be continuous, the former condition can be enforced by just forcing $\xi$ to be infinity on $B$. 
More precisely, if a function~$\tilde{\xi}$~is a good candidate for the reaction coordinate but does not satisfy the previous condition \eqref{cond-xi}, it is possible to obtain $\xi$ from $\tilde{\xi}$ by setting:
\begin{equation}
\xi(\mathbf{X})=\left\{
\begin{array}{ll}
\tilde{\xi}(\mathbf{X})& \mathbf{X} \in \mathbb{R}^{d\times d} \setminus B \\
\infty& \mathbf{X} \in B.
\end{array}
\right.
\end{equation}
The condition \eqref{cond-xi} is then satisfied with $z_{max}$ equal to the maximum value of $\tilde{\xi}$ outside $B$.

We will focus on the estimation of the probability to observe a reaction trajectory, that is, coming from a set of initial conditions in $\mathbb{R}^{d\times d} \setminus (A \cup B)$, the probability to enter $B$ before returning to $A$.
Let us call $\tau_A$ and~$\tau_B$~the first hitting times of $A$ and $B$, respectively (see equations~\eqref{def-tauA} and \eqref{def-tauB} below).
What we aim to calculate is then the probability ${\mathbb{P}(\tau_{B}<\tau_{A})}$.
As will be explained bellow, this probability can be used to compute transition times.
As mentioned earlier, AMS also yields a consistent ensemble of reactive trajectories (this will be illustrated in Section~\ref{section:numerical}).

A detailed description of the AMS algorithm is given in Section \ref{section:algo}.
In Section~\ref{section:properties}~we present a brief discussion of some interesting features of the method.
From the probability obtained using an appropriate set of initial conditions, the transition time can also be computed.
This is explained in Section~\ref{section:eqtime}.

\subsection{The AMS algorithm}
\label{section:algo}

\begin{figure*}[htb!]
\centering
\includegraphics[width=0.98\linewidth]{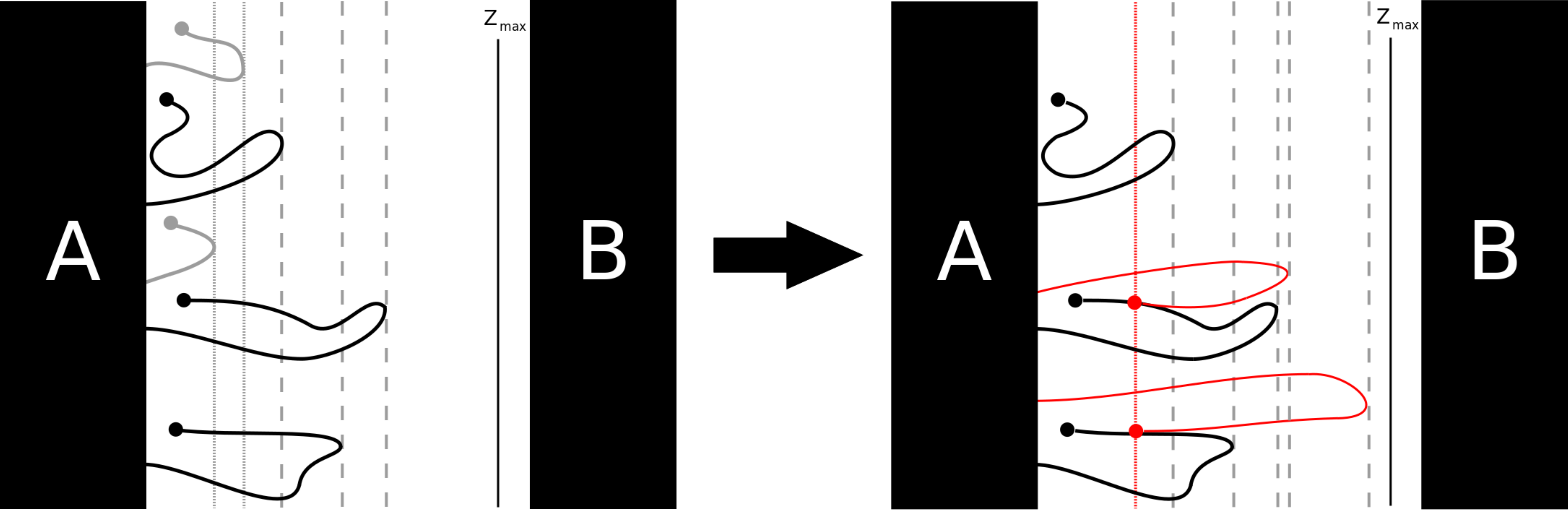}
\caption{
First AMS iteration with $N=5$ and $k=2$.
Both lower level replicas (in gray) are killed. 
Two of the remaining replicas are randomly selected to be duplicated until level $z_{kill}^0$ (dotted red line) and then continued until they reach $A$ (typically more likely) or $B$.
}
\label{algo}

\end{figure*}

The three numerical parameters of the algorithm are: the reaction coordinate $\xi$, the total number of replicas~$N$, and the minimum number $k$ of replicas killed at each iteration.
Let us denote by $\mathbf{X}_t^{n,q}$ the vector of positions and momenta at time $t$ of the $n^{\text th}$ replica ($1\leq n \leq N$) at iteration $q$ of the AMS algorithm.
Let us now consider a set of initial conditions $(\mathbf{X}_0^{n,0})_{1\leq n \leq N}$, which are i.i.d. random variables distributed according to a distribution~$\mu_0$~over~$\mathbb{R}^{d\times d}$, supported outside but in a neighborhood of $A$.  
For all $n \in \{1,...,N\}$ the path from~$\mathbf{X}_0^{n,0}$~to either $A$ or $B$ is computed, creating the first set of replicas~$(\mathbf{X}_{t\in\left[0,\tau_{AB}^{n,0}\right]}^{n,0})_{1\leq n \leq N}$, where~${\tau_{AB}^{n,0}=\min(\tau_A^{n,0},\tau_B^{n,0})}$~with:
\begin{equation}
\tau_{A}^{n,0} = \inf \left\{t \geq 0 : \mathbf{X}_t^{n,0} \in A\right\}
\label{def-tauA}
\end{equation}
and
\begin{equation}
\tau_{B}^{n,0} = \inf \left\{t \geq 0 : \mathbf{X}_t^{n,0} \in B\right\}.
\label{def-tauB}
\end{equation}
So $\tau_{AB}^{n,0}$ is the first time that the $n^{\text th}$ replica at iteration~$q=0$ enters $A$ or $B$.
In this initialization step, since the trajectories start in a neighborhood of $A$, they enter $A$ before $B$ with a probability very close to one.
Notice that the replica $\mathbf{X}_{t\in\left[0,\tau_{AB}^{n,0}\right]}^{n,0}$ reaches $B$ if and only if $\tau_{B}^{n,0}<\tau_{A}^{n,0}$.
Let us denote by $(w_{n,0})_{1\leq n \leq N}$ the weight of each replica, that is initialized as $1/N$:
\begin{equation}
\forall \quad 1 \leq n \leq N, w_{n,0}=\frac{1}{N}.
\label{rep_weights_initial}
\end{equation}
The algorithm then consists of iterating over $q\geq 0$ the three following steps:

\begin{enumerate}
\item Computation of the killing level.\\
At the beginning of iteration $q$ the set of replicas is~$(\mathbf{X}_{t\in[0,\tau_{AB}^{n,q}]}^{n,q})_{1\leq n \leq N}$.
Let us note by $z_n^q$ the highest achieved value of the reaction coordinate by the~$n^{\text th}$~replica:
\begin{equation}
z_n^q=\text{sup}\left\{ \xi(\mathbf{X}_t^{n,q}) : 0\leq t \leq \tau_{AB}^{n,q}\right\}.
\end{equation}
This is called the level of the replica.
To compute the killing level, the replicas are ordered according to their level.
Hence, let us introduce the permutation $\alpha^q: [1,N] \rightarrow [1,N]$ of the trajectories' labels such that:
\begin{equation}
z_{\alpha^q(1)}^q \leq z_{\alpha^q(2)}^q \leq ... \leq z_{\alpha^q(N)}^q.
\end{equation}
The killing level is defined as the k$^{th}$ order level, i.e.~$z_{kill}^q=z_{\alpha^q(k)}^q$.
If all the replicas have a level lower or equal to the killing level one sets~${z_{kill}^q = +\infty}$. 

\item Stopping criterion.\\
The algorithm stops at iteration $q$ if $z_{kill}^q > z_{max}$.
This happens if all the replicas reached the last level~$z_{max}$~or if~$z_{kill}^q = +\infty$, a situation called extinction in the following. 
When the stopping criterion is satisfied, the algorithm is stopped and the current iteration index $q$ is stored in a variable called $Q_{iter}$. 
Notice that $Q_{iter}$ may be null, since $q$ starts from zero.
The integer $Q_{iter}$ is exactly the number of replication steps (see step 3 below) that have been performed when the algorithm stops.

\item Replication.\\
All the $k^{q+1}$ replicas for which $z_n^q \leq z_{kill}^q$ are killed.
Notice that $k^{q+1} \in \{k, k+1,...,N-1\}$. 
Among the~$N-k^{q+1}$~remaining replicas, $k^{q+1}$ are uniformly chosen at random to be replicated.
Replication consists in copying the replica up to the first time it goes beyond the level $z_{kill}^q$, so the last copied point has a level strictly larger than $z_{kill}^q$. 
From that point, the dynamics is run until $A$ or $B$ is reached.
This will generate $k^{q+1}$ new trajectories with level larger than $z_{kill}^q$.
Once all the killed replicas have been replaced, the new set of replicas $(\mathbf{X}_{t\in[0,\tau_{AB}^{n,q+1}]}^{n,q+1})_{1\leq n \leq N}$~is defined.
To complete iteration $q$ one has to update the new weights by:
\begin{equation}
\forall \quad 1 \leq n \leq N , w_{n,q+1}=\frac{N-k^{q+1}}{N}w_{n,q}.
\label{rep_weights}
\end{equation}
From this, $q$ is incremented by one and one comes back to the first step to start a new iteration.
\end{enumerate}

Let us consider the set of all $M$ replicas $\mathbf{X}_{t\in\left[0,\tau_{AB}^{m}\right]}^{m}$ generated during the algorithm run, including the killed ones, and call $w_m$ their weight.
The estimator of $\mathbb{E}(F(\mathbf{X}_{t\in [0,\tau_{AB}]}))$, for any path functional $F$ is\cite{AMS-unbias}
\begin{equation}
\sum_{m=1}^M w_m F(\mathbf{X}_{t\in [0,\tau_{AB}^m]}^m).
\label{eq:any_ams}
\end{equation}
This will be used in Section~\ref{section:committor} to compute the committor function over the phase space.

Note from the description of the algorithm that, at a giving iteration, all the living replicas have the same weight.
The weight of a killed replica stops being updated after it is killed.
Therefore, the replica weight depends on up to which iteration it has survived.

As previously mentioned, we will be particularly interested in the estimation of the probability~${\mathbb{P}(\tau_B < \tau_A)}$, which corresponds to the choice of the path functional $\mathds{1}_{\tau_{B}<\tau_{A}} (\mathbf{X}_{t\in [0,\tau_{AB}]})$ in \eqref{eq:any_ams}.
This means that only the trajectories that survived until the end of the algorithm run will be taken into account.
Therefore, using condition \eqref{cond-xi} and Equation \eqref{eq:any_ams}:
\begin{equation}
p_{AMS}=\sum_{n=1}^N w_{n,Q_{iter}} \mathds{1}_{\tau_{B}^{n,Q_{iter}}< \tau_{A}^{n,Q_{iter}}}
\label{eq:pams-weight}
\end{equation}
is an estimator of $\mathbb{P}(\tau_B < \tau_A)$.
Here the weights are all equal.
Using Equations \eqref{rep_weights_initial} and \eqref{rep_weights}, and denoting by~$r$~the number of replicas that reached $B$ at the last iteration of the algorithm, $p_{AMS}$ can be rewritten as
\begin{equation}
p_{AMS}=\frac{r}{N}\prod_{q=0}^{Q_{iter}-1}\left(\frac{N-k^{q+1}}{N}\right),
\label{eq:pams}
\end{equation}
where by convention $\prod\limits_{q=0}^{-1}=1$.
To gain intuition in this formula, notice that the term $\frac{N-k^{q+1}}{N}$ in Equation \eqref{eq:pams} is an estimation of the probability of reaching level $z_{kill}^{q}$, conditioned to the fact that level $z_{kill}^{q-1}$ has been reached, (where by convention $z_{kill}^{-1}=-\infty$).
Also, as an example, if all the replicas in the initial set $(\mathbf{X}_{t\in\left[0,\tau_{AB}^{n,0}\right]}^{n,0})_{1\leq n \leq N}$ reached $B$, $r=N$ and thus $p_{AMS}=1$.
In case of extinction $r=0$, because no replica reached $B$, and thus $p_{AMS}=0$.

Note that the number $k^{q+1}$ of killed replicas at iteration $q$ may exceed $k$. 
The situation were $k^{q+1}>k$ happens if there is more than one replica with level equal to~$z_{kill}^q$.
There are typically two situations for which this occurs.
First, this may happen if there exists a region where the reaction coordinate is constant.
Second, it may be a consequence of the replication step at a previous iteration if the following occurs: 
(1) The point up to which the replica is copied has a $\xi$-value which is the maximum of the $\xi$-values along the trajectory (namely the level of the replica);
(2) The replicated replica has the same level as the copied replica.
Notice that this happens because the AMS method is applied to a discrete in time Markov process.

This algorithm is implemented in NAMD \cite{NAMD} as a Tcl script, easily used via the configuration file.
The script is compatible with NAMD version 2.10 or higher\cite{ams-namd-tutorial}.
In order to decrease the computational cost, the reaction coordinate of a point in the trajectory is only calculated every~${K_{\text {\tiny AMS}}=\Delta t_{\text {\tiny AMS}}/\Delta t}$~timesteps.
This means that, in practice, the algorithm is actually applied to the subsampled Markov chain $(\mathbf{X}_{s K_{\text {\tiny AMS}}})_{ s \in \mathbb{N}}$.
It is indeed useless to consider the positions of the trajectory at each simulation time step, as no significant change occurs in a $1$ or $2$ fs time scale.
Also notice that, along a trajectory, only the points that can possibly be used in future replication steps must be recorded, reducing memory use.
This corresponds to points for which the reaction coordinate strictly increases.



\subsection{Properties of the AMS method}
\label{section:properties}

Let us recall some important properties of the AMS method obtained in previous works.
One of them is the unbiasedness of the algorithm. 
It can be proven\cite{AMS-unbias} that the expected value of the probability estimator is equal to the probability to be calculated:
\begin{equation}
\mathbb{E}(p_{AMS})=\mathbb{P}(\tau_B<\tau_A).
\label{eq:unbias}
\end{equation}
This is more generally true for the estimator \eqref{eq:any_ams}:
\begin{equation}
\mathbb{E}\left( \sum_{m=1}^M w_m F(\mathbf{X}_{t\in [0,\tau_{AB}^m]}^m) \right) = \mathbb{E}(F(\mathbf{X}_{t\in [0,\tau_{AB}]})).
\end{equation}
Hence, in practice, the algorithm is run more than once and the result is obtained as an empirical average of the estimators for each run.
This also provides naturally asymptotic confidence interval on the results, using the central limit theorem. 
Notice that unbiasedness holds whatever the choice of the reaction coordinate $\xi$, the number of replicas $N$ and the minimum number of killed replicas~$k$~at each iteration. 
Therefore, one can compare the results obtained with different sets of parameters (in particular different reaction coordinates) to gain confidence in the result.
These parameters however affect the variance of the estimator and, consequently, its efficiency. 

In another paper\cite{AMS-ideal}, one considers the ideal case, namely the situation where the reaction coordinate is the committor function.
It can be proven that this is the best reaction coordinate in terms of the variance of $p_{AMS}$.
Moreover, this ideal case is interesting since explicit computations give some insights on the efficiency of the algorithm, that are observed to be useful beyond the ideal case.
In the ideal case, variance and the efficiency of the method are then proportional to~$1/N$.
Let us recall that the efficiency of a Monte Carlo method can be defined as the inverse of the product of the computational cost and the variance\cite{hammersley1964monte}.
The number of iterations $Q_{iter}$ is a random variable that follows a Poisson distribution with mean value $-N\log(\mathbb{P}(\tau_B<\tau_A))$. 
This indicates that the method is well suited to estimate small probabilities, hence appropriate to the simulation of rare events. 

We concentrated here on the estimation of the probability $\mathbb{P}(\tau_B<\tau_A)$, but as explained above, see \eqref{eq:any_ams}, other estimations can be made with this method\cite{AMS-unbias}. 
It is possible, for example, to calculate unbiased estimators of $\mathbb{E}(F((\mathbf{X}_{t\in [0,\tau_{AB}]}))\mathds{1}_{\tau_B<\tau_A})$~for any path functional $F$ by simply making averages over the trajectories obtained at the end of the algorithm that reached $B$ before $A$. 
Consequently, it is also possible to obtain estimators of conditional expectations~$\mathbb{E}(F((\mathbf{X}_{t\in [0,\tau_{AB}]}))|\tau_B<\tau_A)$.
Such estimators have a bias of order $1/N$ in the large $N$ limit.
This will be used in particular in Section~\ref{section:numerical} to compute the flux of reactive trajectories from $A$ to $B$.

\subsection{The transition time equation}
\label{section:eqtime}

Another quantity that we aim to obtain is the transition time from $A$ to $B$, using the probability estimated by AMS.
The transition time is the average time of the trajectories, coming from $B$, from its first entrance in~$A$~until the first entrance in $B$ afterwards\cite{lunolen,transition-path-theory}.
As~$A$~is metastable, the dynamics makes in and out of~$A$~loops before visiting $B$.
To correctly define those loops let us fix an intermediate value $z_{min}$ of the reaction coordinate, defining an isolevel surface $\Sigma_{z_{min}}$:
\begin{equation}
\Sigma_{z_{min}}=\{\mathbf{X} \in \mathbb{R}^{d\times d} : \xi(\mathbf{X})=z_{min}\}.
\label{sigma_z}
\end{equation}

If $A$ is metastable and $\Sigma_{z_{min}}$ is close to $A$ the number of loops made between $A$ and $\Sigma_{z_{min}}$ before visiting~$B$~is large.
After some of them, the system reaches an equilibrium.
When this equilibrium is reached the first hits of~$\Sigma_{z_{min}}$ follow a so-called quasi-stationary distribution~$\mu_{QSD}$.
Here, we call the first hitting points of~$\Sigma_{z_{min}}$~the first points that, coming from $A$, have a~${\xi\text{-value}}$ larger than~$z_{min}$.
If one then uses as a set of initial conditions the random variables $(\mathbf{X}^{n,0}_0)_{1\leq n\leq N}$ distributed according to $\mu_{QSD}$, it is possible to evaluate the probability $p$ to reach $B$ before $A$ starting from~$\Sigma_{z_{min}}$~at equilibrium by using AMS.
As $A$ is metastable, the number of loops needed to reach the equilibrium is small compared to the total number of loops made before going to~$B$, so it can be neglected.

\begin{figure}[htb!]
\centering
\includegraphics[width=0.97\linewidth]{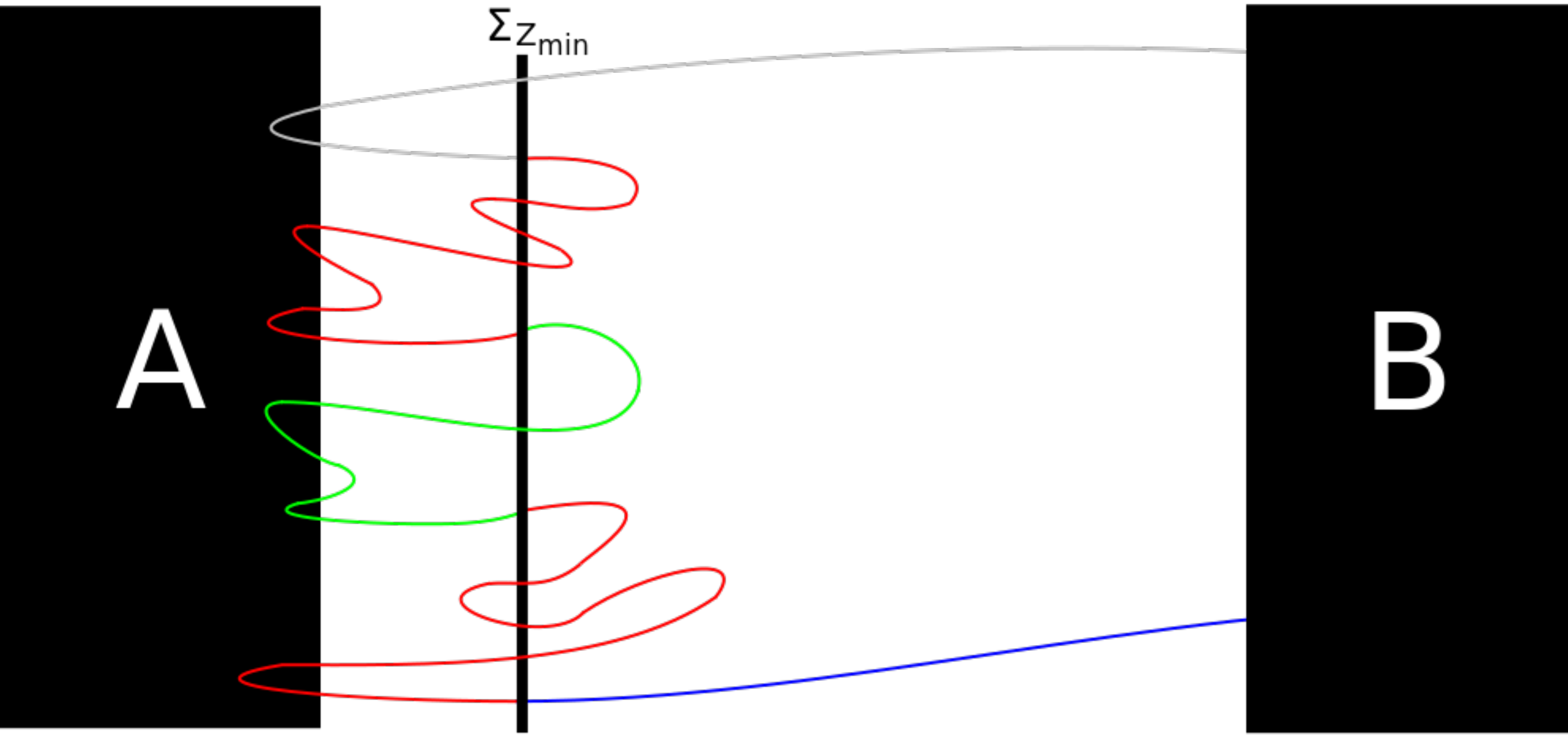}
\caption{
The loops between $A$ and $\Sigma_{z_{min}}$ (green and red) and the reaction trajectory (blue).
}
\label{tempsdraw}
\end{figure}

Let us now use these considerations to estimate the transition time from A to B.
Consider an equilibrium trajectory coming from $B$ that enters $A$ and returns to $B$. 
The goal is to calculate the average time ($\mathbb{E}(T_{AB})$) of this trajectory\cite{transition-path-theory}.
A good strategy is to split this path in two: the loops between $A$ and $\Sigma_{z_{min}}$, and the reaction trajectory, i.e. the path from $A$ to $B$ that does not comes back to~$A$~after reaching $\Sigma_{z_{min}}$.
This is outlined in Figure~\ref{tempsdraw}.
Neglecting the first time taken to go out of $A$, one can define as $T_{loop}^k$ the time of the~$k^{\text th}$~loop between two subsequent hits of $\Sigma_{z_{min}}$, conditioned to have visited $A$ between them, and as~$T_{reac}$~the time of the reaction trajectory. 
If the number of loops made before visiting $B$ is $n$, the time $T_{AB}$ can be obtained as:
\begin{equation}
T_{AB}=\sum_{k=1}^{n}T_{loop}^k+T_{reac}.
\end{equation}
At each passage over $\Sigma_{z_{min}}$ there are two possible events, first enter $A$ or first enter $B$. 
As mentioned in the previous paragraph, it is possible to obtain with AMS the probability $p$ at equilibrium to visit $B$ before $A$ starting from the probability distribution $\mu_{QSD}$ on $\Sigma_{z_{min}}$.
Therefore, the waiting time to enter~$B$~is~$1/p$, so the mean number of loops made before that is~$1/p-1$.
This leads us to the final equation for the expected value of $T_{AB}$:
\begin{equation}
\mathbb{E}(T_{AB})=\left(\frac{1}{p}-1\right)\mathbb{E}(T_{loop})+\mathbb{E}(T_{reac}).
\label{temps_trans}
\end{equation}

The mathematical formalization of this reasoning is a work in progress.
The consistency of \eqref{temps_trans} has already been tested on various systems in previous works\cite{AMS-benzamidine,temps-trans}.
In this paper, we numerically investigate the quality of formula~\eqref{temps_trans} using the estimate of~$p$~obtained with AMS starting from~$\mu_{QSD}$~(see Section~\ref{section:time}).
Note that the sampling of~$\mu_{QSD}$ as well as $\mathbb{E}(T_{loop})$ can be obtained with short direct simulations while AMS is used to get both $p$ and $\mathbb{E}(T_{reac})$. 
The first term in Equation \eqref{temps_trans} is much larger than the last one in the case of a rare event, making crucial the achievement of good probability estimations to obtain good estimations for the transition time.
Typically, the term $\mathbb{E}(T_{reac})$ is small compared to $\mathbb{E}(T_{AB})$ and can be ignored.
In fact, other methods\cite{FFS-2009,WE-1998} like forward flux sampling and weighted ensemble approximate the reaction rate $k_{AB}=\mathbb{E}(T_{AB})^{-1}$ by $p/\mathbb{E}(T_{loop})$, which is consistent with our formula \eqref{temps_trans}.

Choosing the parameter $z_{min}$ may be delicate. 
The closer $\Sigma_{z_{min}}$ to $A$, the smaller the probability $p$ to estimate. 
On the other hand, if $\Sigma_{z_{min}}$ is too far from $A$, there will be fewer loops, so the time to reach the quasi-stationary distribution will not be negligible. 
Moreover, the simulation time needed to obtain a good estimation of $T_{loop}$ will be larger. 
This will again be discussed in the numerical example in the next section.
\section{Numerical results}
\label{section:numerical}

We apply the AMS method to the $C_{eq} \rightarrow C_{ax}$ transition of the N-acetyl-N’-methylalanylamide, also known as alanine dipeptide or dialanine.
The transition between its two stable conformations in gas phase occurs in a time scale of the order of a hundred nanoseconds, allowing us to obtain direct numerical simulation (DNS) estimations to compare to results obtained with AMS. 

\begin{figure}[htb!]
\centering
\includegraphics[width=0.7\linewidth,trim={0 0.5cm 0 1cm},clip]{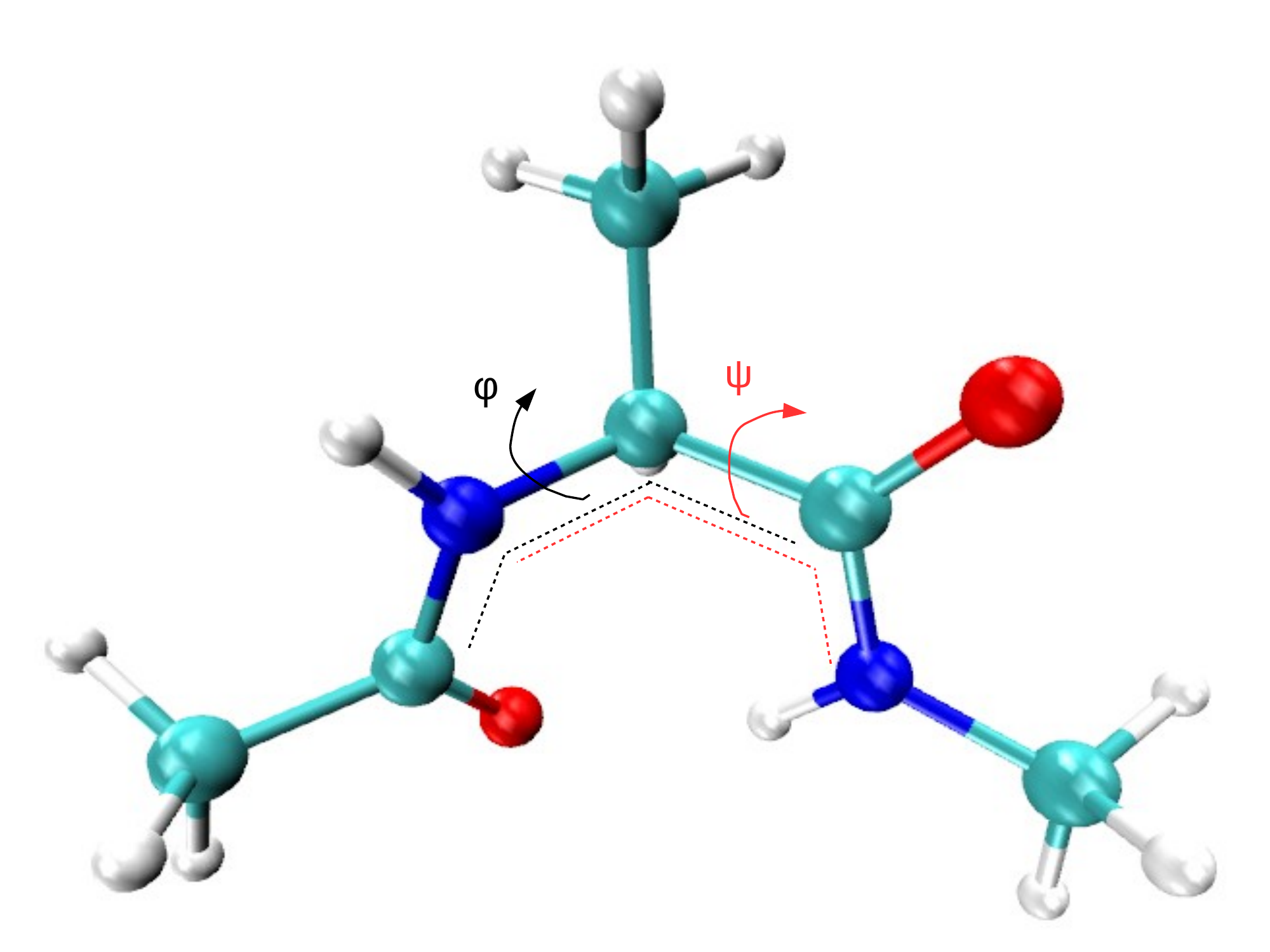}
\caption{
The dihedral angles~$\varphi$~and~$\psi$~used to distinguish between the~$C_{eq}$~and~$C_{ax}$~conformations.
}
\label{dialafig}
\end{figure}

\begin{figure}[htb!]
\centering
\includegraphics[width=0.8\linewidth,trim={0.6cm 2cm 0 3.3cm},clip]{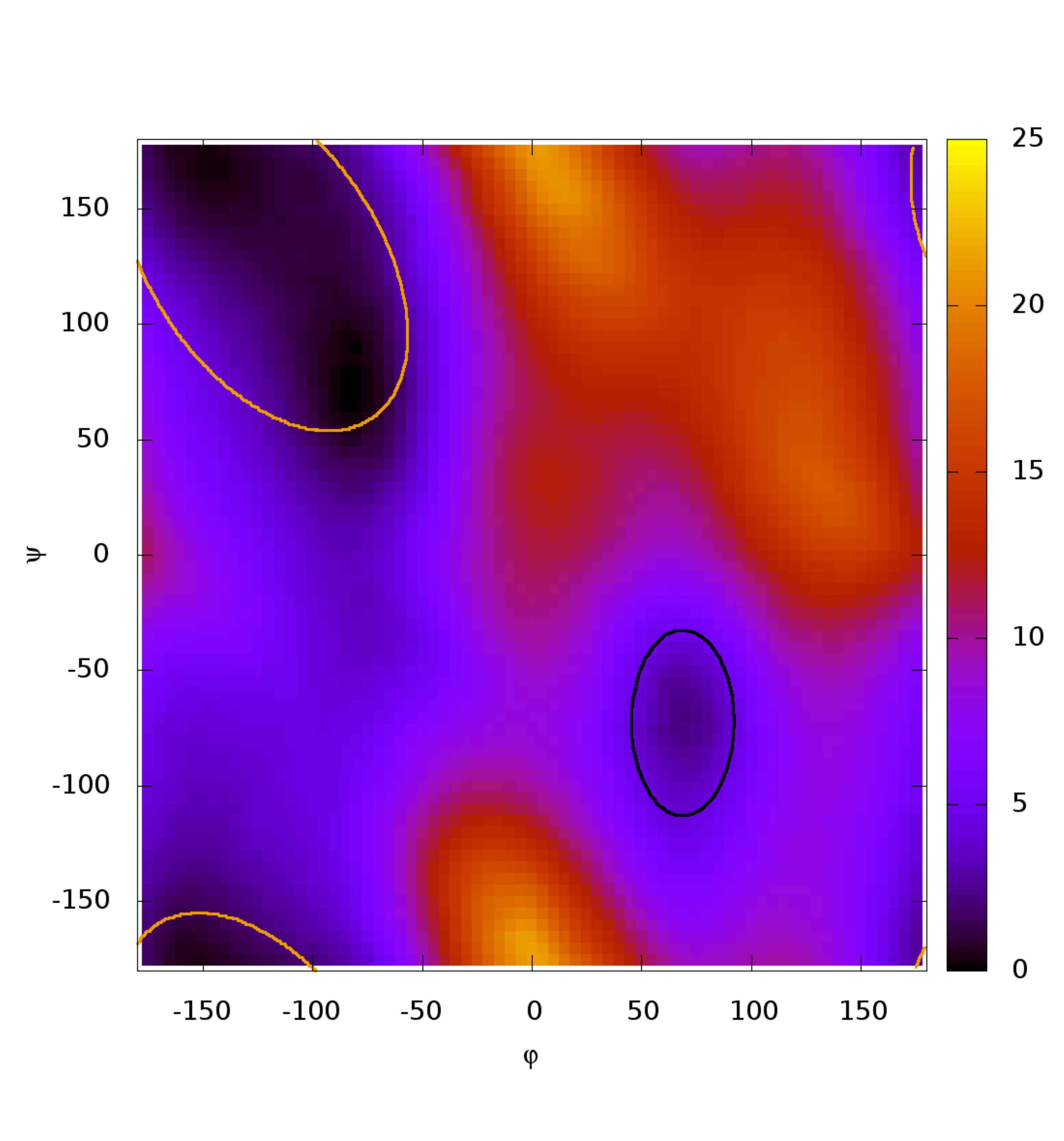}
\caption{
The free energy landscape \cite{dialanine-free-energy} with the definition of zones $A$ (yellow) and $B$ (black).
}
\label{free_reg}
\end{figure}

Both conformations can be characterized by two dihedral angles, $\varphi$ and $\psi$ (Figure \ref{dialafig}).
Regions~$A$~and~$B$~($C_{eq}$~and~$C_{ax}$, respectively), are defined as ellipses that covers the two most significant wells on the free energy landscape (Figure \ref{free_reg}).

\begin{figure}[htb!]
\centering
\includegraphics[width=0.8\linewidth,trim={2cm 3.5cm 0 5.5cm},clip]{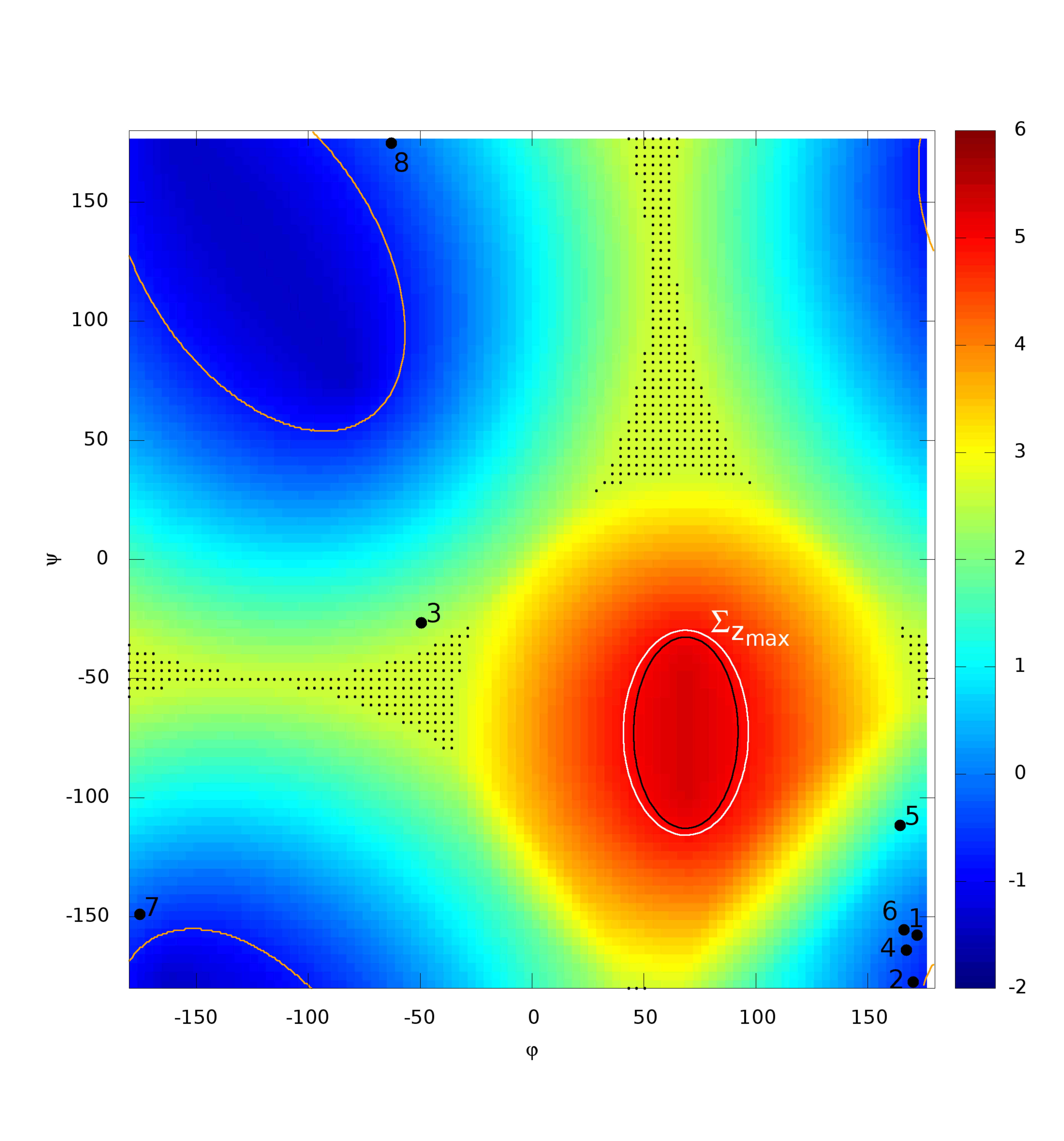}
\caption{
Contour plot of the second reaction coordinate $\xi_2$. 
Regions $A$ and $B$ are marked in yellow and black, respectively.
The region $\Sigma_{z_{max}}$ used for the AMS runs ($z_{max}=4.9$) is marked in white.
The zone covered with black dots corresponds to regions where $\xi_2$ is constant and equal to $2.6$.
}
\label{cri}
\end{figure}

Two reaction coordinates are investigated.
The first one (see \eqref{crlinear}) is a continuous piecewise affine function of~$\varphi$~and the second one (see~\eqref{crellip}) is a measure of the distance to the two regions~$A$~and~$B$.
Here are the precise definitions of $\xi_1$ and $\xi_2$ (see Figure \ref{cri} for a contour plot of $\xi_2$):
\small
\begin{equation}
\xi_1(\varphi)=\left\{
\begin{array}{l l}
-5.25 & \text{if }\varphi < -52.5 \\
0.1\varphi & \text{if }-52.5 \leq \varphi \leq 45 \\
4.5 & \text{if }45 < \varphi < 92.5 \\
-0.122\varphi + 15.773 & \text{if } 92.5 \leq \varphi \leq 172.5 \\
-5.25 & \text{if } \varphi > 172.5 \\
\end{array}
\right.
\label{crlinear}
\end{equation}
\vspace{-0.2cm}
\begin{equation}
\xi_2(\varphi,\psi)=\min(d_A,6.4)-\min(d_B,3.8)
\label{crellip}
\end{equation}
\normalsize\\
In Equation~\eqref{crellip},~$d_A$~(resp.~$d_B$) is the sum of the Euclidean distances to the foci of the ellipse~$A$~(resp.~$B$).\\

The values of $z_{max}$ used for the simulations are $4.49$ for~$\xi_1$~and $4.9$ for $\xi_2$.
All the simulations are performed using NAMD\cite{NAMD} version 2.11 with the CHARMM27 force field. 

To numerically illustrate some properties of the algorithm, we first calculate the transition probability starting from one fixed (deterministic) initial condition.
These results are presented in Section~\ref{section:point}, as well as the flux of reaction trajectories obtained with different initial conditions.
The estimations of transition times are reported in Section~\ref{section:time}, where a proper way to sample~$\mu_{QSD}$ is proposed.
Finally, we present in Section~\ref{section:committor} a way to use AMS in order to compute an approximation of the committor function.

\subsection{Calculating the Probability with AMS}
\label{section:point}

\begin{figure}[htb!]
\centering
\includegraphics[width=0.85\linewidth,trim={0 0.1cm 0 0.5cm},clip]{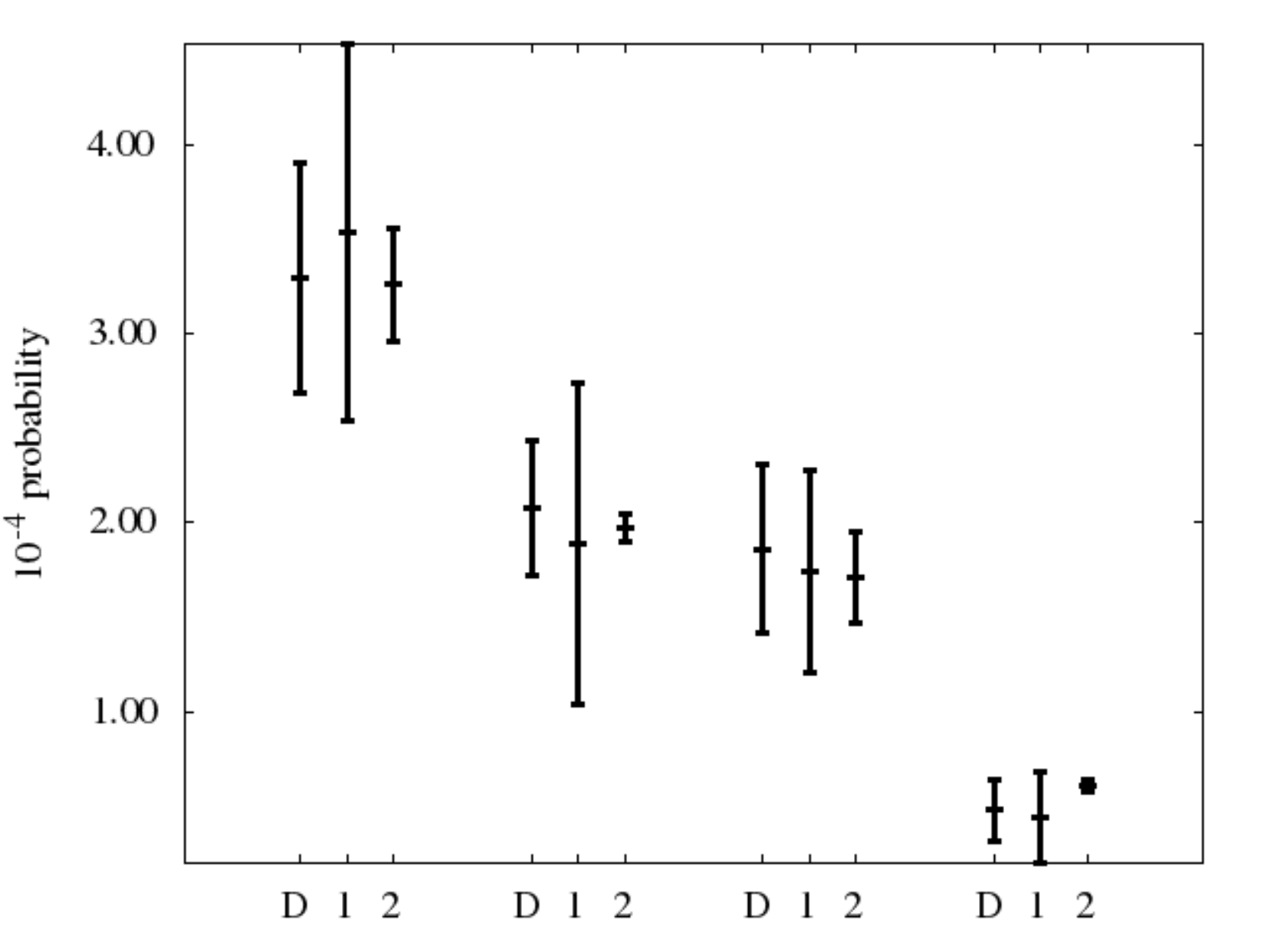}
\caption{
Probability estimations using different points as a initial condition: D is for DNS, 1 is for AMS using $\xi_1$ and 2 is for AMS using $\xi_2$.
For each point we made about 200 AMS runs and a 15 ns DNS.
}
\label{proba_pontos}
\end{figure}

To evaluate the efficiency of the algorithm to estimate the probability to visit $B$ before $A$, we first initiate all the replicas from the same point $\mathbf{x}$ (fixed positions and velocities for all atoms), i.e. $\forall n \in [1,N]$, $\mathbf{X}_0^{n,0}=\mathbf{x}$. 
This enables us to compare estimates of the probability to enter $B$ before $A$ obtained with AMS with accurate values obtained using DNS.
In DNS, simulations start from $\mathbf{x}$ and stop when $A$ or $B$ is reached.
The ratio of the number of times $B$ is reached over the total number of simulations is the DNS estimation for the probability $\mathbb{P}(\tau_B<\tau_A)$.
Results (both for DNS and AMS) are reported in Figure~\ref{proba_pontos} for four different choices of~$\mathbf{x}$~(points 1 to 4 in Figure~\ref{cri}).

First note from Figure~\ref{proba_pontos} the robustness of the AMS algorithm with respect to the choice of the reaction coordinate. 
The two reaction coordinates indeed give probability estimates in accordance with the direct simulation values. 
The second interesting feature is the change in the confidence interval, that tends to be smaller for $\xi_2$. 
This illustrates the fact that the average of the estimator is the same whatever the choice of $\xi$ (see~\eqref{eq:unbias}), but the variance depends on~$\xi$.
\begin{figure}[htb!]
\centering
\includegraphics[width=0.95\linewidth]{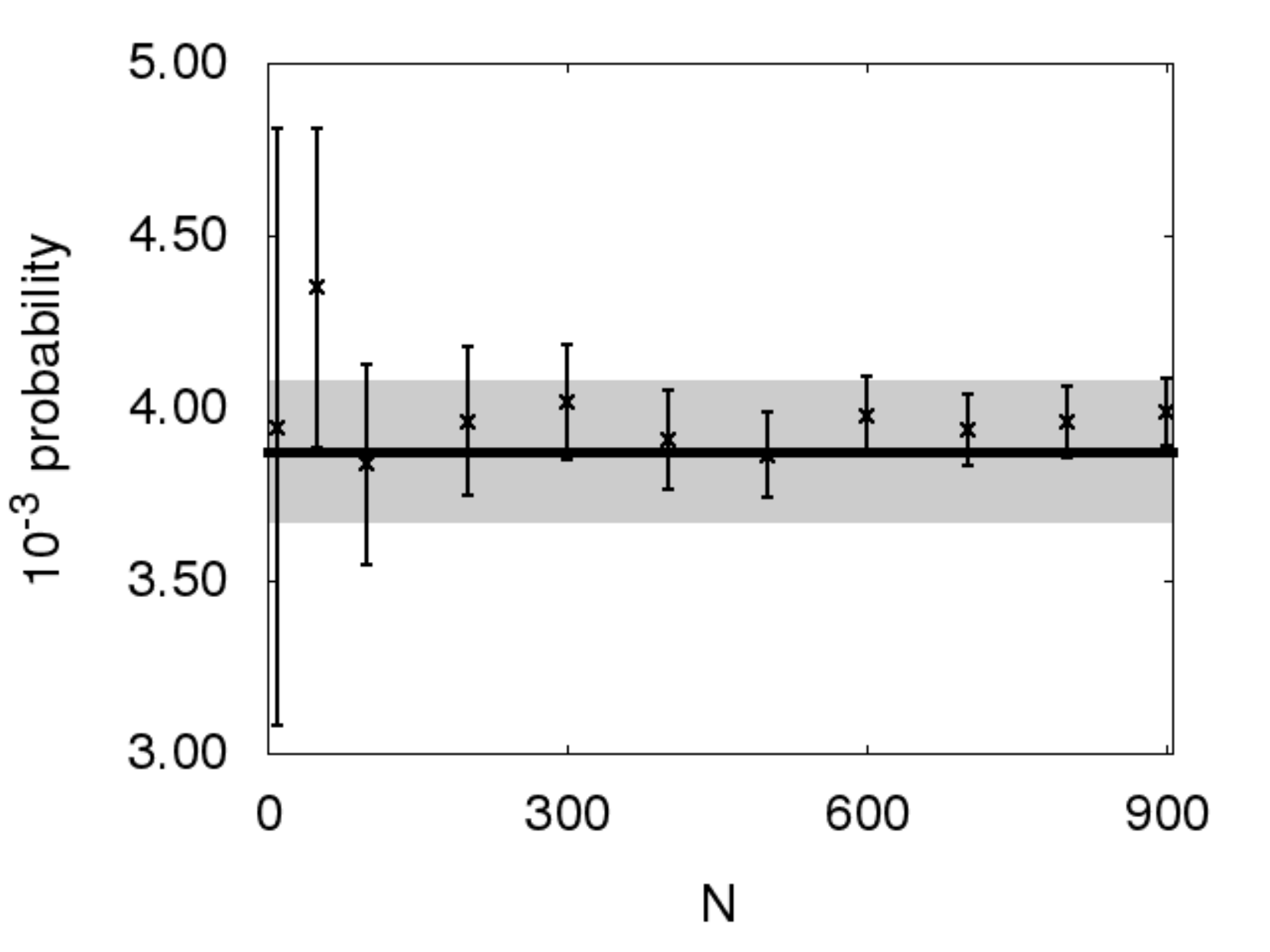}
\includegraphics[width=0.95\linewidth]{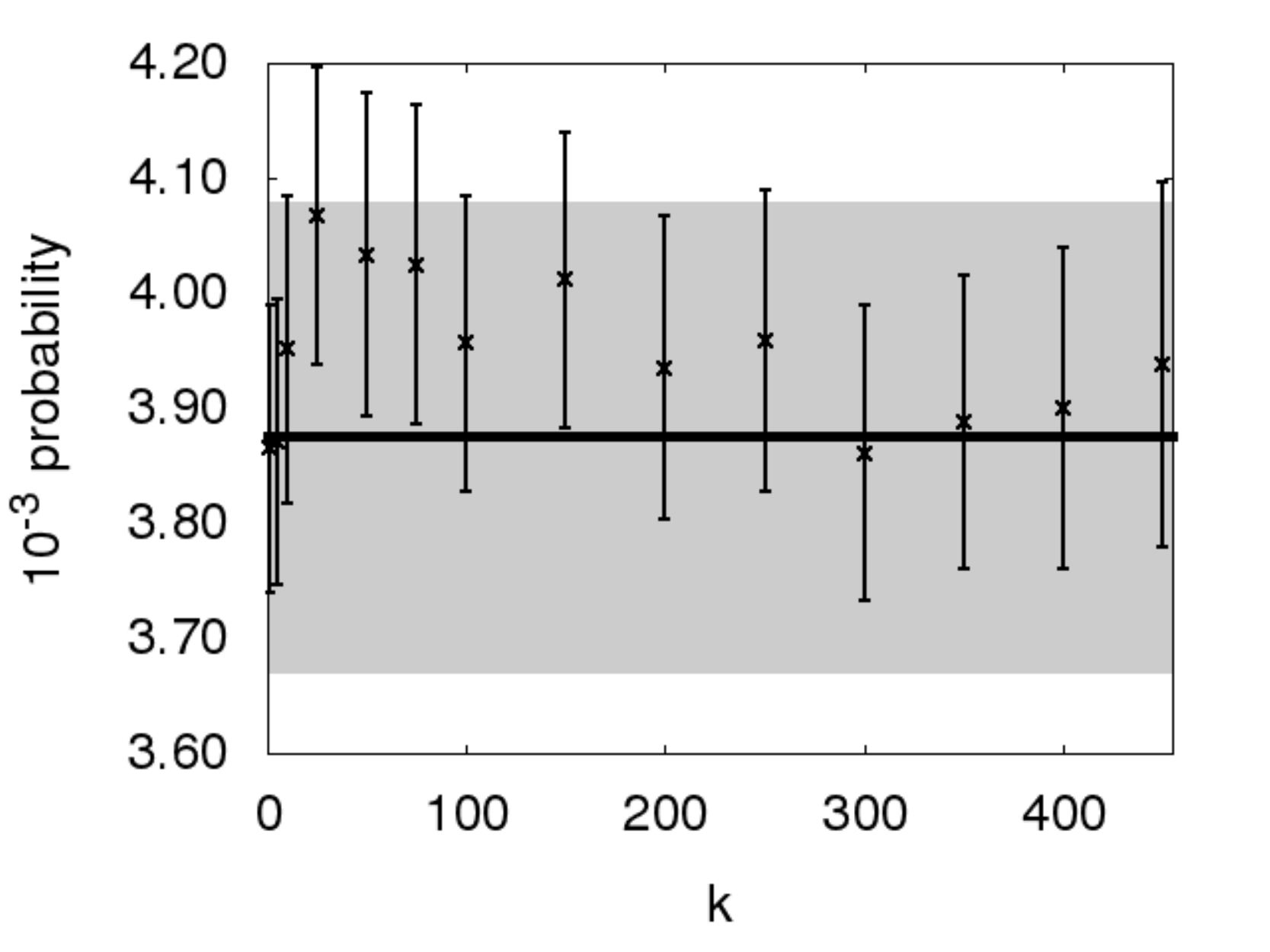}
\caption{
AMS estimations for the probability with different values of $k$ and $N$.
Results were obtained using a fixed initial condition (point 1 in Figure \ref{cri}) with $\xi_2$ and 1000 AMS runs for each value of $N$ and $k$.
\vspace{-0.5cm}
}
\label{proba_repkill}
\end{figure}

Notice from results in Figure~\ref{proba_repkill} that different values of~$k$~and~$N$~yield consistent estimates of the probability.
This is again a numerical illustration of~\eqref{eq:unbias}.
Notice that the variance scales as $1/N$, as already discussed in Section \ref{section:properties}. 

\begin{figure}[htb!]
\centering
\includegraphics[width=0.95\linewidth,trim={0 0.7cm 0 0.3cm},clip]{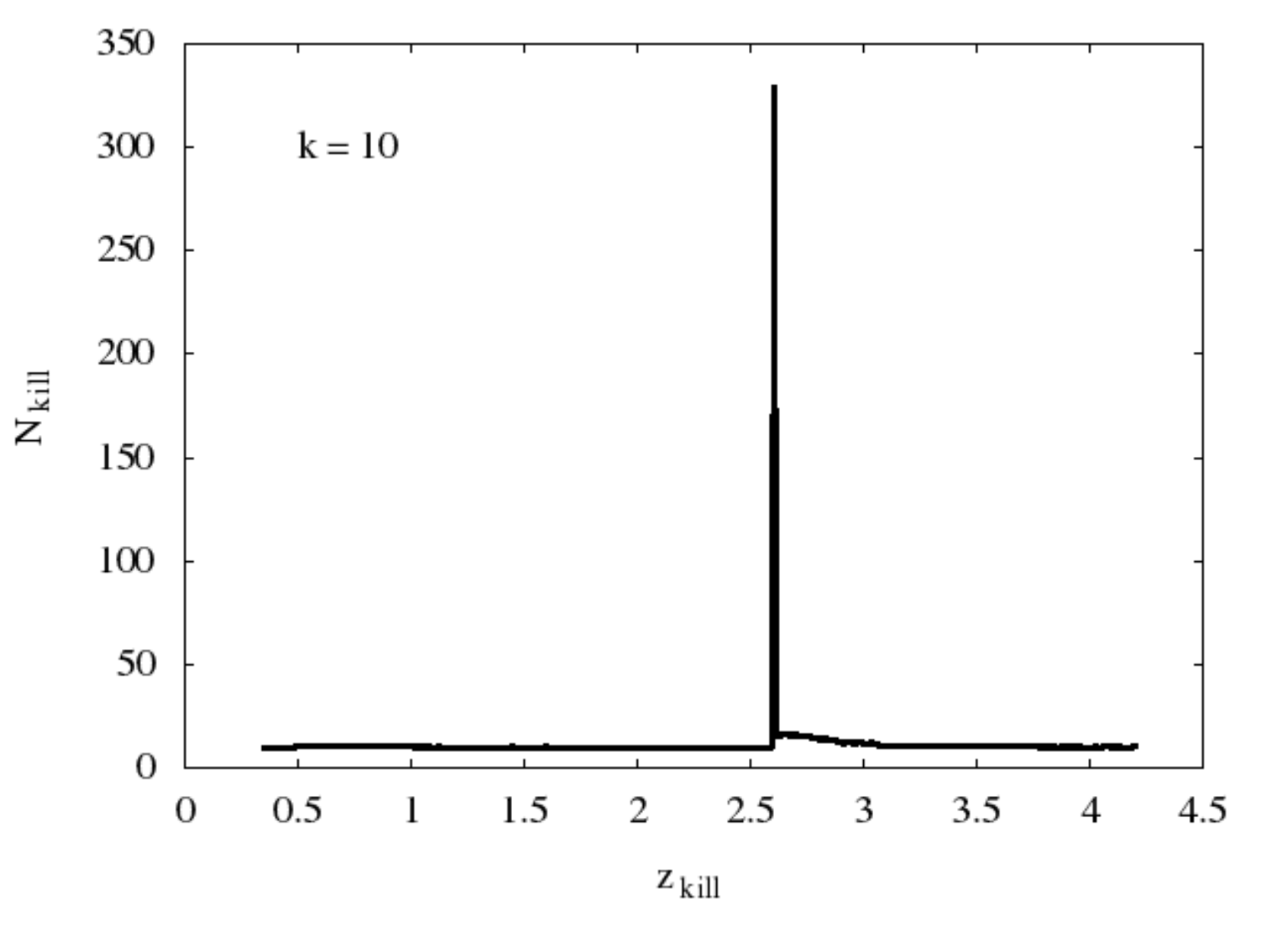}
\caption{
Variation of the number of replicas killed as a function of the killing level.
This graph was obtained with a mean over 1000 AMS runs.
}
\label{kill}
\end{figure}

\begin{figure*}[th!]
\centering
\includegraphics[width=\linewidth,trim={0 0 0 2cm},clip]{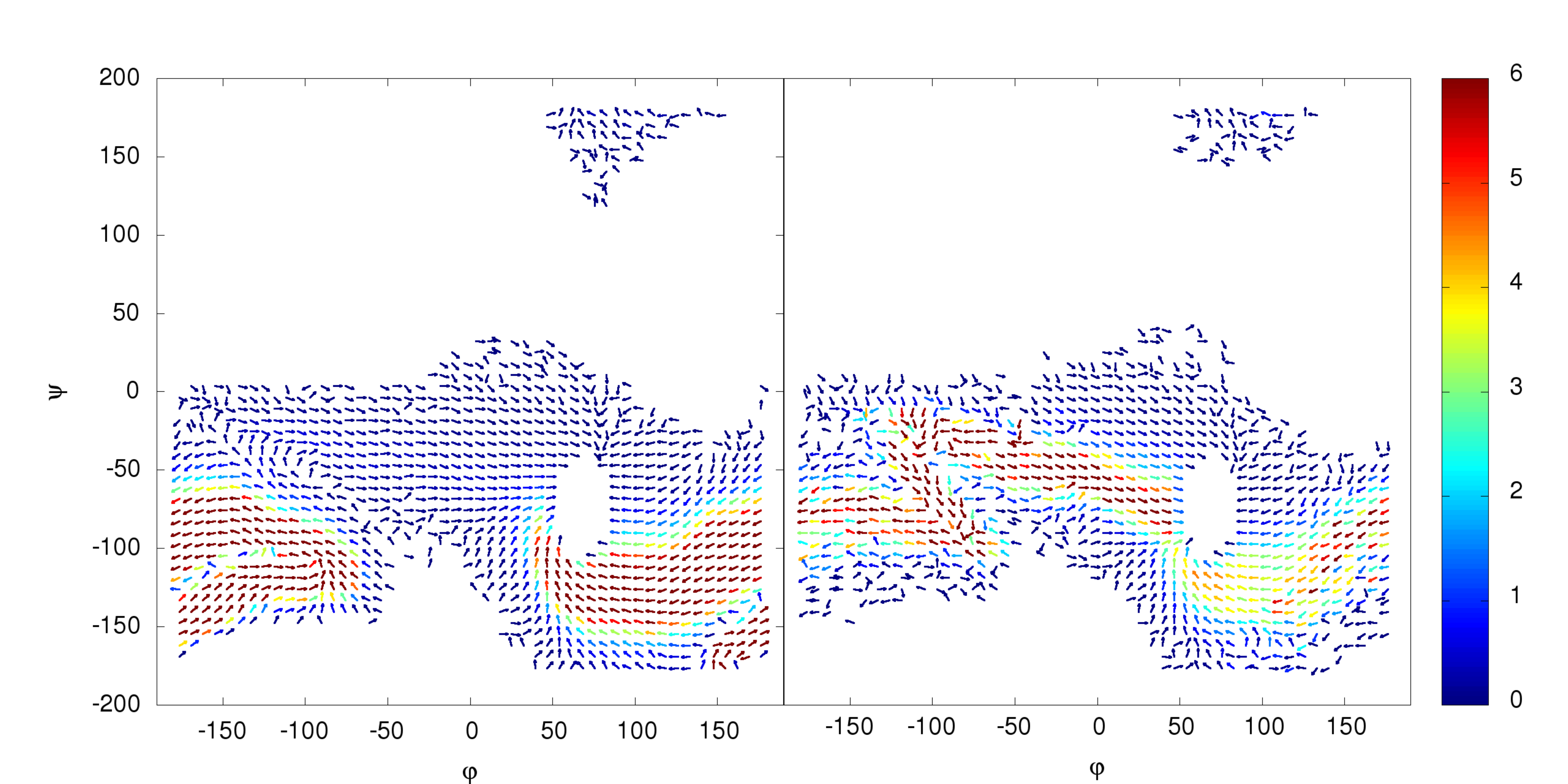}
\caption{
Flux for points 2 and 3 (see Figure \ref{cri}) obtained with 500,000 trajectories, results of 1000 AMS runs with 500 replicas each.
}
\label{flux}
\end{figure*}

\vspace{10cm}
Another interesting fact can be illustrated looking at the number of killed replicas at each killing level ($z^q_{kill}$) over the AMS runs with the reaction coordinate $\xi_2$ (Figure \ref{kill}).
The number of replicas is close to $k$ for all levels except for $\xi_2=2.6$, which is the value of the reaction coordinate in regions where it is constant (see Figure \ref{cri}).
This implies that a large number of replicas are at the same level when exploring these regions. 
So, at the stage where $z_{kill}=2.6$, all replicas in this level are killed, which explains this result. 
This phenomenon increases the possibility of getting zero as an estimator of the probability, thus increases the variance.
It is important to note that, even with such a locally constant reaction coordinate, $\xi_2$~exhibits good results with low variances, showing again that the AMS algorithm is robust in terms of the choice of the reaction coordinate.

To obtain information on the reaction paths and thus on the reaction mechanism, the flux of the reaction trajectories is evaluated by a numerical approximation based on the following formula (see\cite{transition-path-theory} and Remark 1.13\cite{lunolen}):
\begin{equation}
J(x)=\lim_{T\rightarrow \infty} \frac{1}{T} \int_0^T \dot{q}_t \delta (x-q_t) \mathds{1}_R(t) dt,
\label{lunolen_flux}
\end{equation}
where for a given time $t$, $\mathds{1}_R(t)$ is one if $q_t$ belongs to a transition path from $A$ to $B$ and zero otherwise.
Using a set $\{(\mathbf{X}^1_t)_{t\in[0,\tau_B^1]},...,(\mathbf{X}^n_t)_{t\in[0,\tau_B^n]}\}$ of reaction trajectories obtained with the AMS method, each
trajectory~$i$~has a weight of $w_i$ and can be associated with a vector~$(\bm{\theta}^i_t)_{t \in [0,\tau^i_B]}$~where~${(\bm{\theta}^i_t){=}(\varphi(\mathbf{X}^i_t),\psi(\mathbf{X}^i_t))}$ are the two dihedral angles (see Figure \ref{dialafig}). 
The $(\varphi,\psi)$ space is split into $L$ cells $(C_l)_{1\leq l \leq L}$.
The flux in each cell is then defined up to a multiplicative constant by (compare with Equation~\eqref{lunolen_flux}):
\begin{equation}
J(C_l)=\sum\limits_{i=1}^{n}w_i\sum\limits_{t=0}^{\tau^i_B-1} \left(\frac{\bm{\theta}^i_{t+1}-\bm{\theta}^i_{t}}{\Delta t}\right) \mathds{1}_{\bm{\theta}^i_{t} \in C_l}.
\label{eq:flux}
\end{equation}
In Figure \ref{flux}, the fluxes approximated using Equation~\eqref{eq:flux} are represented for two different initial conditions.
Such a result is useful to visualize the transition paths from $A$ to $B$.
These paths highly depend on the initial condition, as can be seen by comparing the two results in Figure \ref{flux}.

\begin{figure}[htb!]
\centering
\includegraphics[width=0.8\linewidth,trim={0 0.2cm 0 0.5cm},clip]{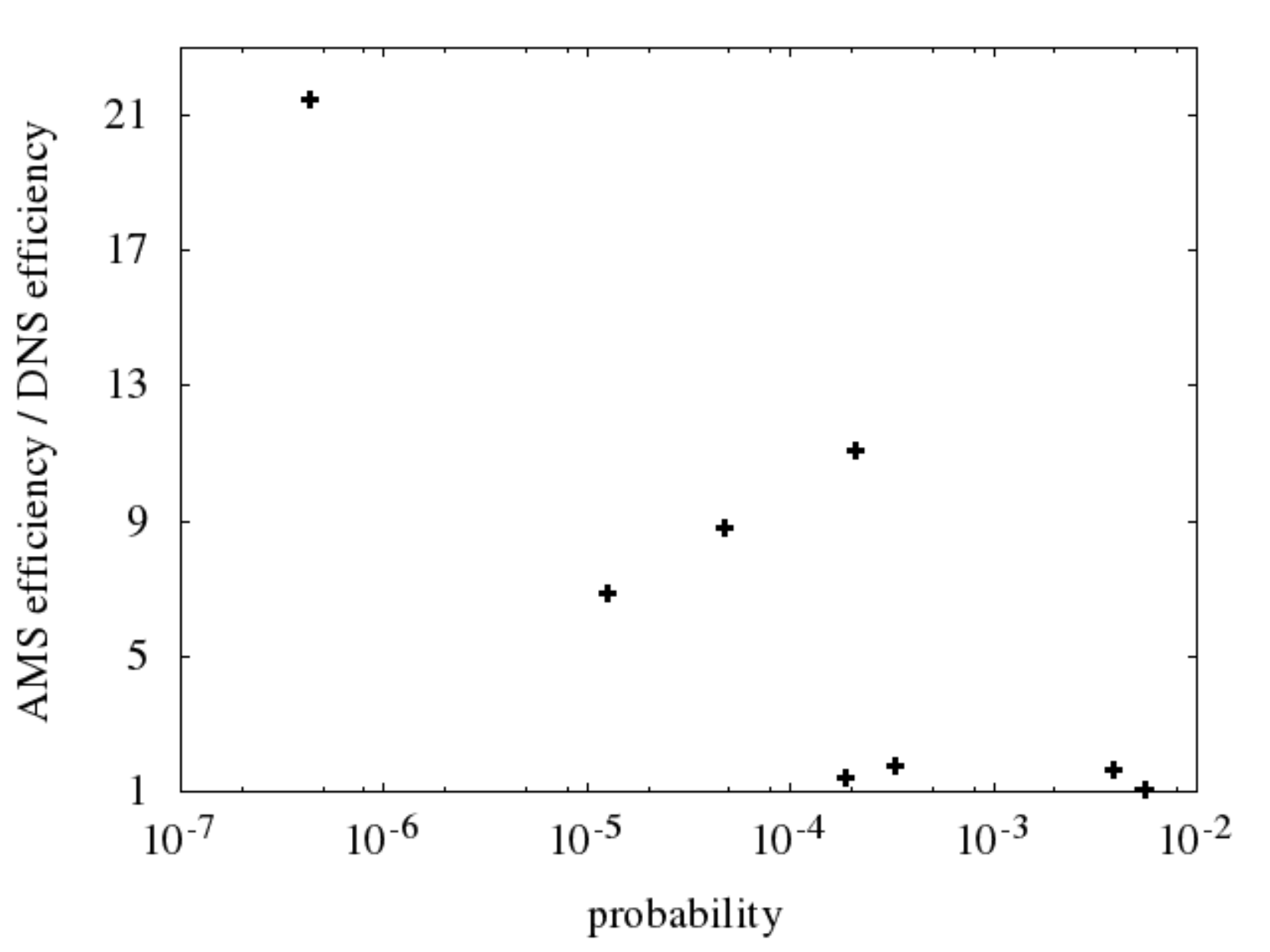}
\caption{
Efficiency ratio between AMS and DNS estimations for points 1 to 8 in Figure \ref{cri}. The confidence intervals are too small to be seen on the graph.
\vspace{-0.2cm}
}
\label{efficiency}
\end{figure}

We also look at the efficiency of the method by applying it to eight initial conditions.
As mentioned in Section~\ref{section:properties}, the efficiency of a Monte Carlo method is defined as the inverse of the product of the computational cost and the variance\cite{hammersley1964monte}. 
In Figure \ref{efficiency} the variation of the ratio of the AMS efficiency over the DNS efficiency as a function of the probability $\mathbb{P}(\tau_B < \tau_A)$ is showed.
When this ratio is larger than 1, the AMS algorithm is more efficient than DNS.
Notice that all the points show that AMS is more efficient than DNS but also that this efficiency tends to be larger when the probability decreases.
This illustrates that the method is particularly well suited to calculate small probabilities. 
As an example, for the point with probability $10^{-7}$ the wall clock time for DNS is over a week, but the estimation with 1000 AMS run in parallel with 32 cores takes less than two days.

\subsection{Calculating the transition time}
\label{section:time}

To evaluate the transition time using Equation \eqref{temps_trans} one needs estimations of $p$, $\mathbb{E}(T_{reac})$ and $\mathbb{E}(T_{loop})$.
The last is easily obtained by a short simulation starting from $A$.
The other two terms can be estimated using AMS, as long as the initial condition's points follow the distribution $\mu_{QSD}$, as mentioned in Section \ref{section:eqtime}.
To obtain a reference value for the transition time, which is ${(309.5\pm23.8)}$~ns, a set of 97 direct simulations of 2$\mu$s each is made.

At first, we make a $2\mu$s simulation, sufficiently long to observe transitions from $A$ to $B$ and thus to obtain DNS estimates for $p$ and $\mathbb{E}(T_{reac})$.
For the probability $p$ we count the number of~$\Sigma_{z_{min}}{\rightarrow}A$~and~$\Sigma_{z_{min}}{\rightarrow}B$~trajectories, respectively~$n_A$~and~$n_B$, yielding the estimate ${p_{DNS}=n_B/(n_A+n_B)}$.
To investigate the consistency of Equation~\eqref{temps_trans}, we also calculate the transition time with these DNS values.

\begin{figure}[htb!]
\centering
\includegraphics[width=0.85\linewidth,trim={0 0.3cm 0 0.3cm},clip]{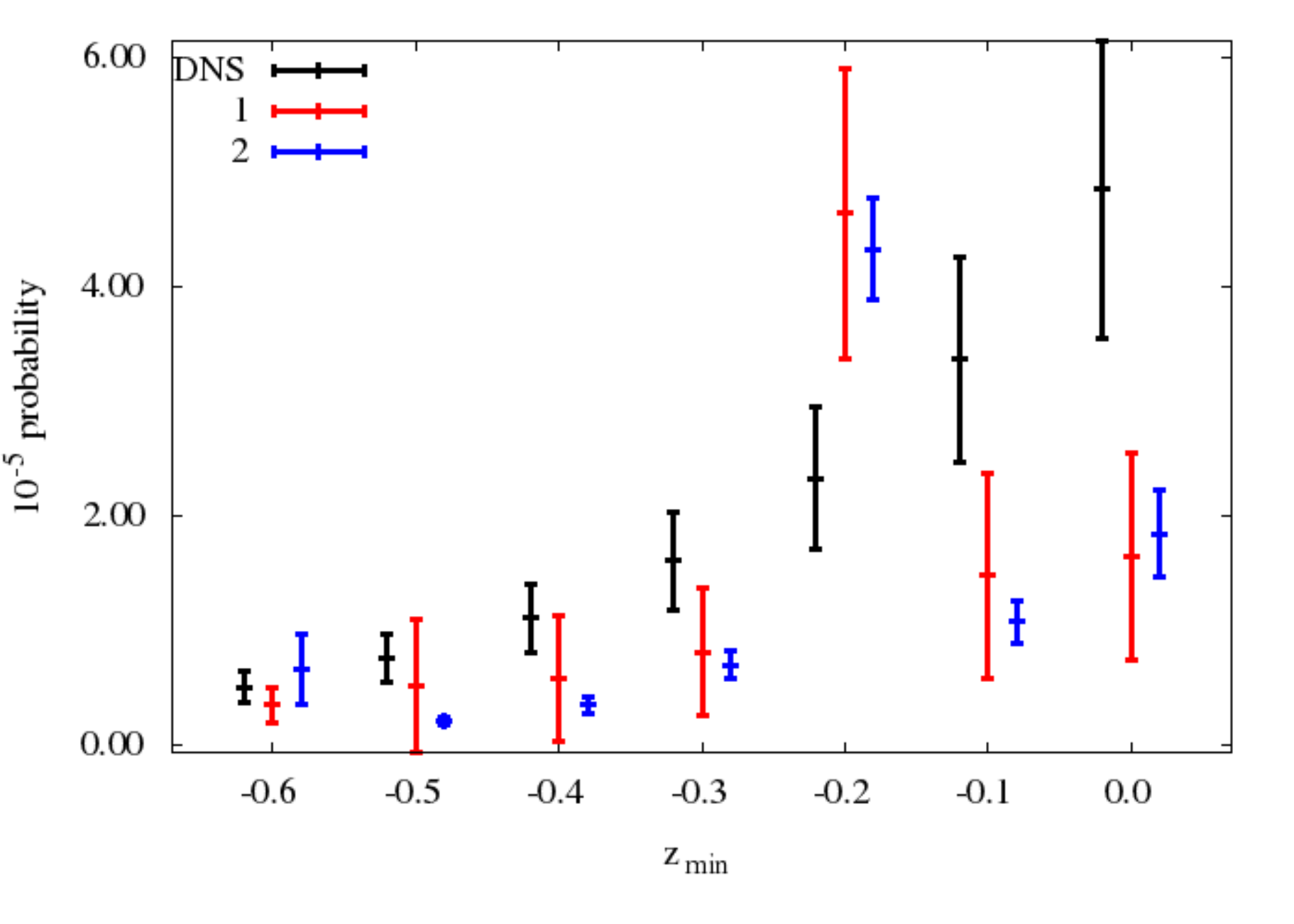}
\includegraphics[width=0.85\linewidth,trim={0 0.3cm 0 0.3cm},clip]{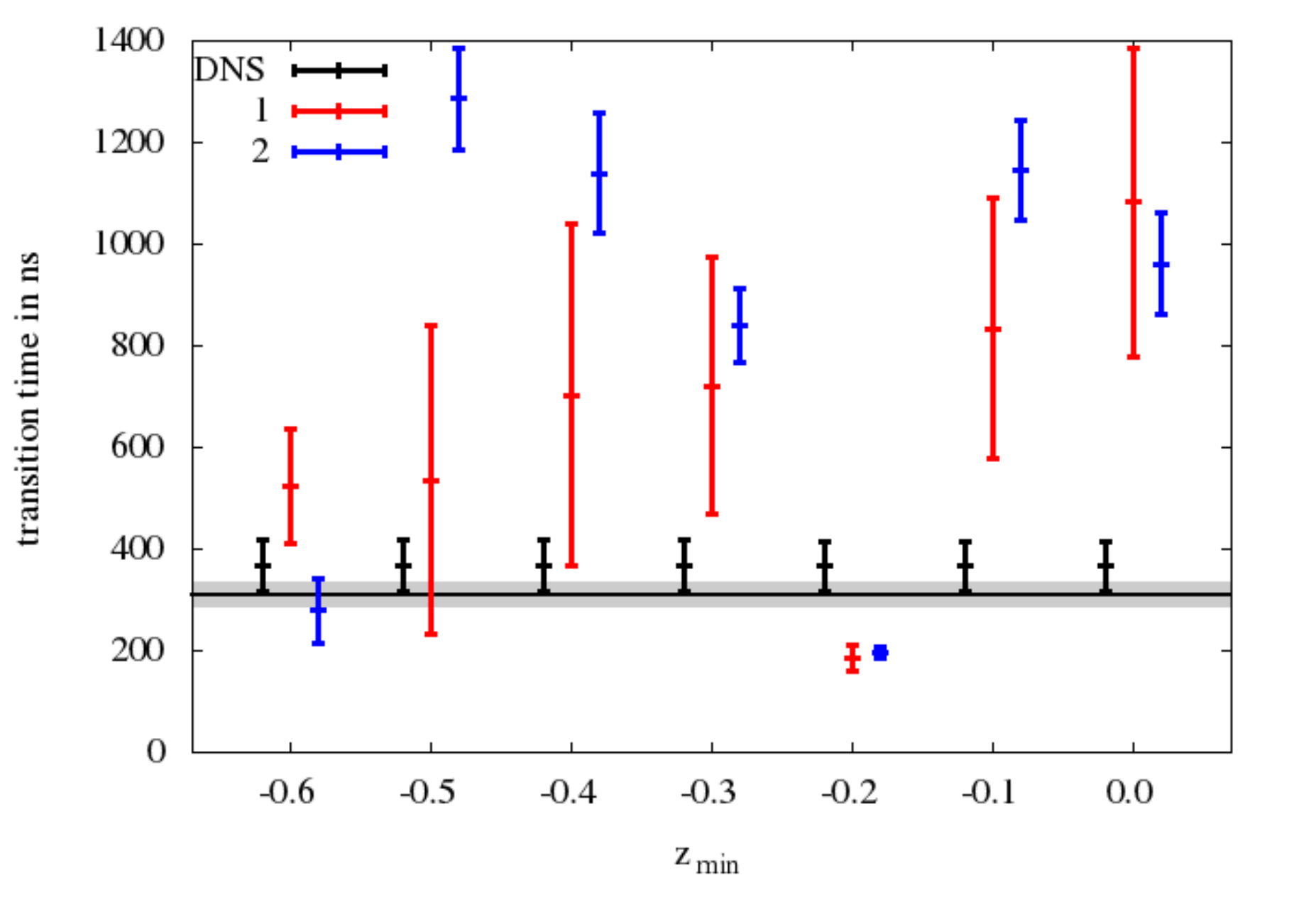}
\caption{
Probability and transition time obtained for the seven sets of initial conditions with DNS and AMS with both $\xi_1$ (1) and $\xi_2$ (2).
The DNS estimations were made using a 2$\mu$s simulation and the AMS with 1000 independent runs.
In the bottom figure the reference value is represented as the gray interval.
}
\label{proba_sets}
\end{figure}

Using the same $2\mu$s simulation, and for a fixed value of $z_{min}$, all the first hitting points of $\Sigma_{z_{min}}$ in the successive loops between $A$ and $\Sigma_{z_{min}}$ are stored and 500 among them are randomly chosen to form the initial conditions' set to run the AMS simulations.
This gives the samples distributed according to $\mu_{QSD}$.
In this process, estimates of $\mathbb{E}(T_{loop})$ are also obtained.
To fix $z_{min}$ we choose to use levels of $\xi_2$ and in total seven different values were adopted.
The obtained results are reported in Figure \ref{proba_sets}. 

Notice from Figure \ref{proba_sets} (bottom) that the transition times obtained with the DNS estimates are consistent with the reference value.
In fact, they only differ by 2 ps one from each other.
This validates the use of Equation~\eqref{temps_trans}.

For the results obtained with AMS, first observe from Figure \ref{proba_sets} (top) the consistency of the probability estimates obtained with the two different reaction coordinates.
For some values of $z_{min}$, these estimations are not consistent with the DNS ones.
Accordingly, for those values of $z_{min}$, the obtained transition times are also not compatible with the reference value, see \ref{proba_sets} (bottom).

\begin{figure}[htb!]
\centering
\includegraphics[width=0.5\linewidth]{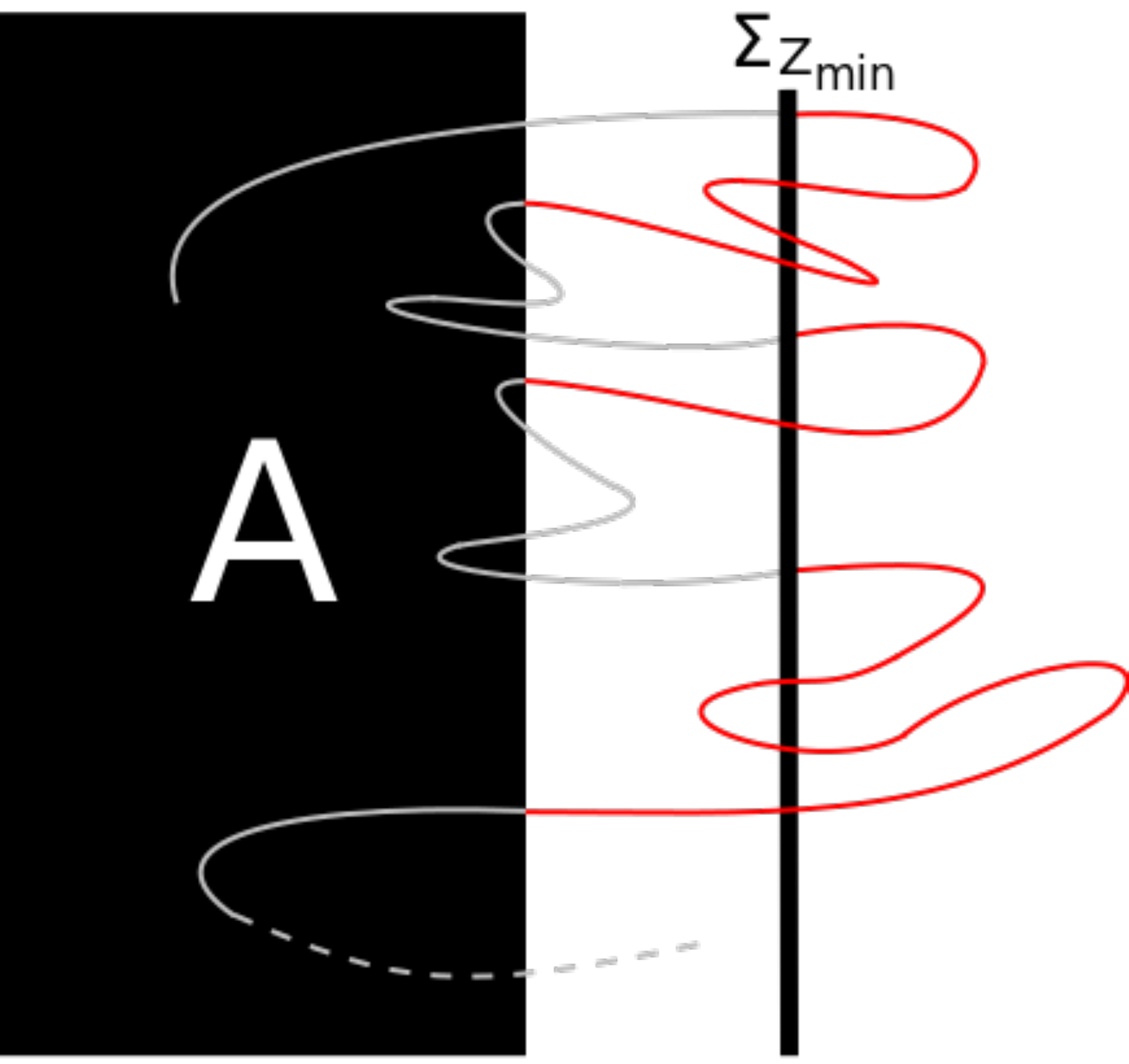}
\caption{The sample of the first 3 initial replicas (in red). The simulation is made until all the 500 replicas are obtained and this process is repeated before each AMS run.}
\label{fig:des-var}
\end{figure}

In order to understand the non consistency between the AMS and the DNS results, we look at the sampling of the initial conditions.
Recall that for AMS, an ensemble of 500 samples is chosen and fixed for all the AMS runs, while for DNS, these are actually sampled along the long trajectory.
Moreover, we observe that the probability to reach~$B$~before~$A$~highly depends on the initial condition in the sample distributed according to~$\mu_{QSD}$.
This yields a result which is not robust with respect to the choice of the 500 initial conditions and raises question about how to efficiently sample $\mu_{QSD}$. 
The strategy we propose is, instead of fixing 500 initial conditions once for all, redraw new ones for each AMS run. 
This is made with a small initial simulation previously to each run, where, starting from $A$, the first 500 $\Sigma_{z_{min}}{\rightarrow}A$ trajectories are used as the first set of replicas (see Figure \ref{fig:des-var}).
This fixes the 500 initial conditions for each run.
Notice that these simulations can also be used to obtain $\mathbb{E}(T_{loop})$, excluding the need to make the initial 2$\mu$s simulation previously mentioned.

\begin{figure}[htb!]
\centering
\includegraphics[width=0.85\linewidth,trim={0 0.3cm 0 0.3cm},clip]{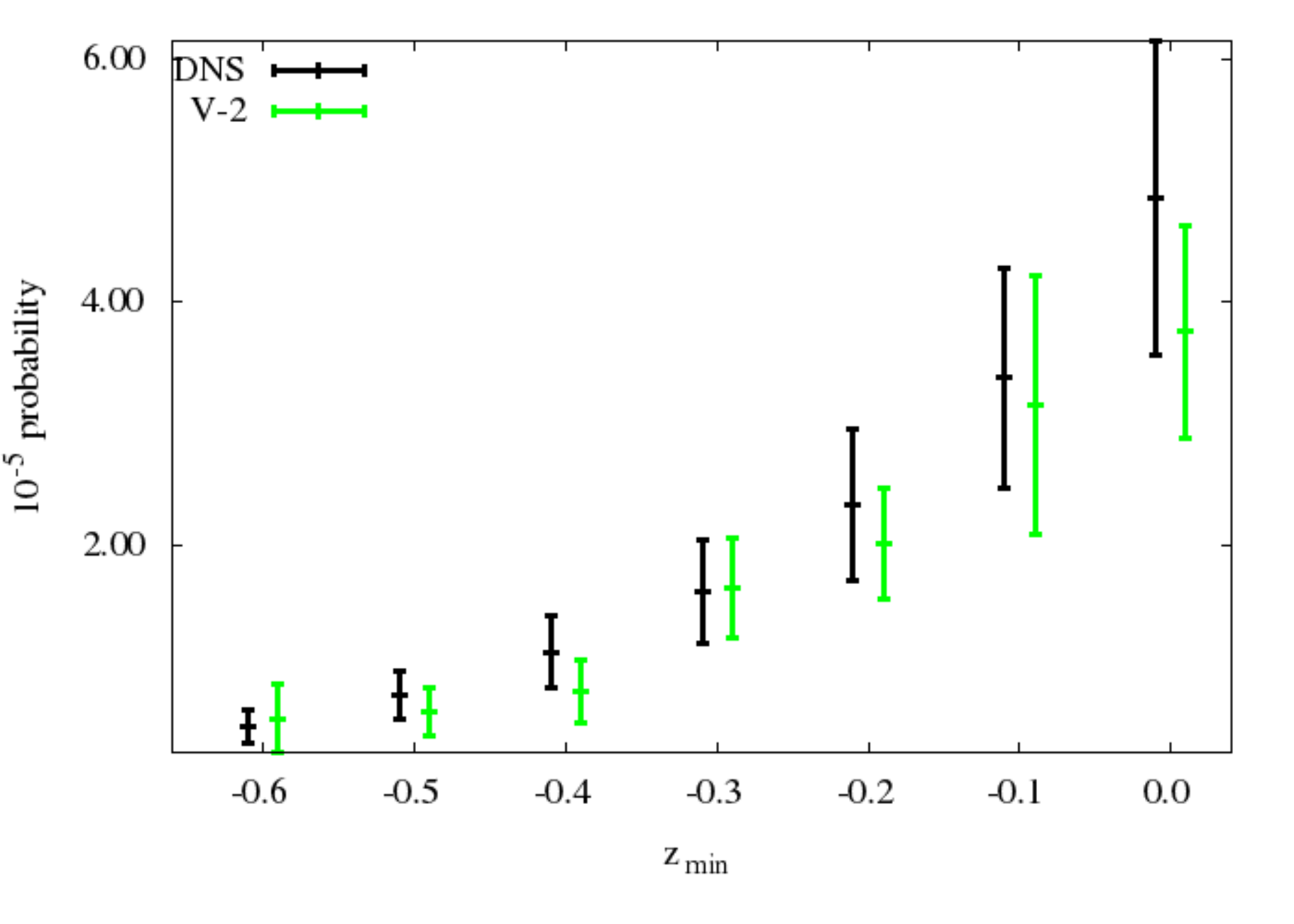}
\includegraphics[width=0.85\linewidth,trim={0 0.3cm 0 0.3cm},clip]{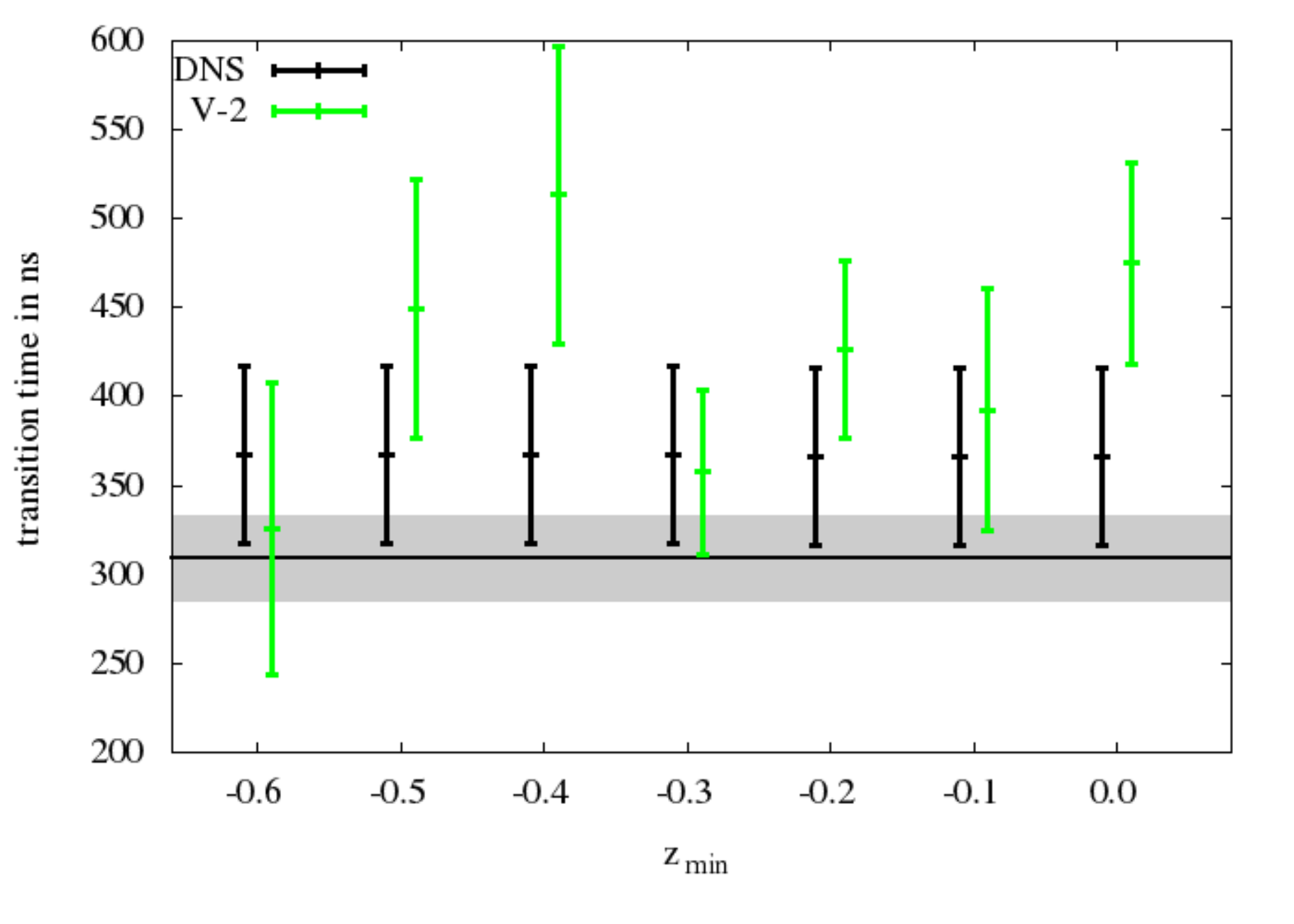}
\caption{
Probability obtained varying the set of initial conditions before each AMS run with $\xi_2$ and the transition time calculated with them.
For each value of $z_{min}$ 1000 AMS runs were made with 500 replicas each.
}
\label{cond_var}
\end{figure}

The results using this new strategy are reported in Figure \ref{cond_var}.
The estimations for the probability, in Figure \ref{cond_var} (top), are in agreement with DNS.
Nevertheless, observe that the larger $z_{min}$, i.e. the far from $A$, the more distant the estimator is from the reference value, and also the larger the variance.
This is because the more far from $A$ the more difficult it is to sample the distribution $\mu_{QSD}$.
Notice that the calculation of the transition time has a term in $1/p$ (see Equation~\eqref{temps_trans}).
Consequently, small errors in the probability causes large errors in the transition time.
This can be observed in Figure \ref{cond_var} (bottom), where the best estimator is for the smaller value of $z_{min}$.
Also notice that the results obtained for the transition time are in better agreement with the reference value than the previous one.
We therefore conclude from this numerical experiment that it is worth redrawing new initial conditions for each AMS simulation in order to better sample the distribution $\mu_{QSD}$.

\begin{figure}[htb!]
\centering
\includegraphics[width=0.9\linewidth,trim={0 0.3cm 0 0.3cm},clip]{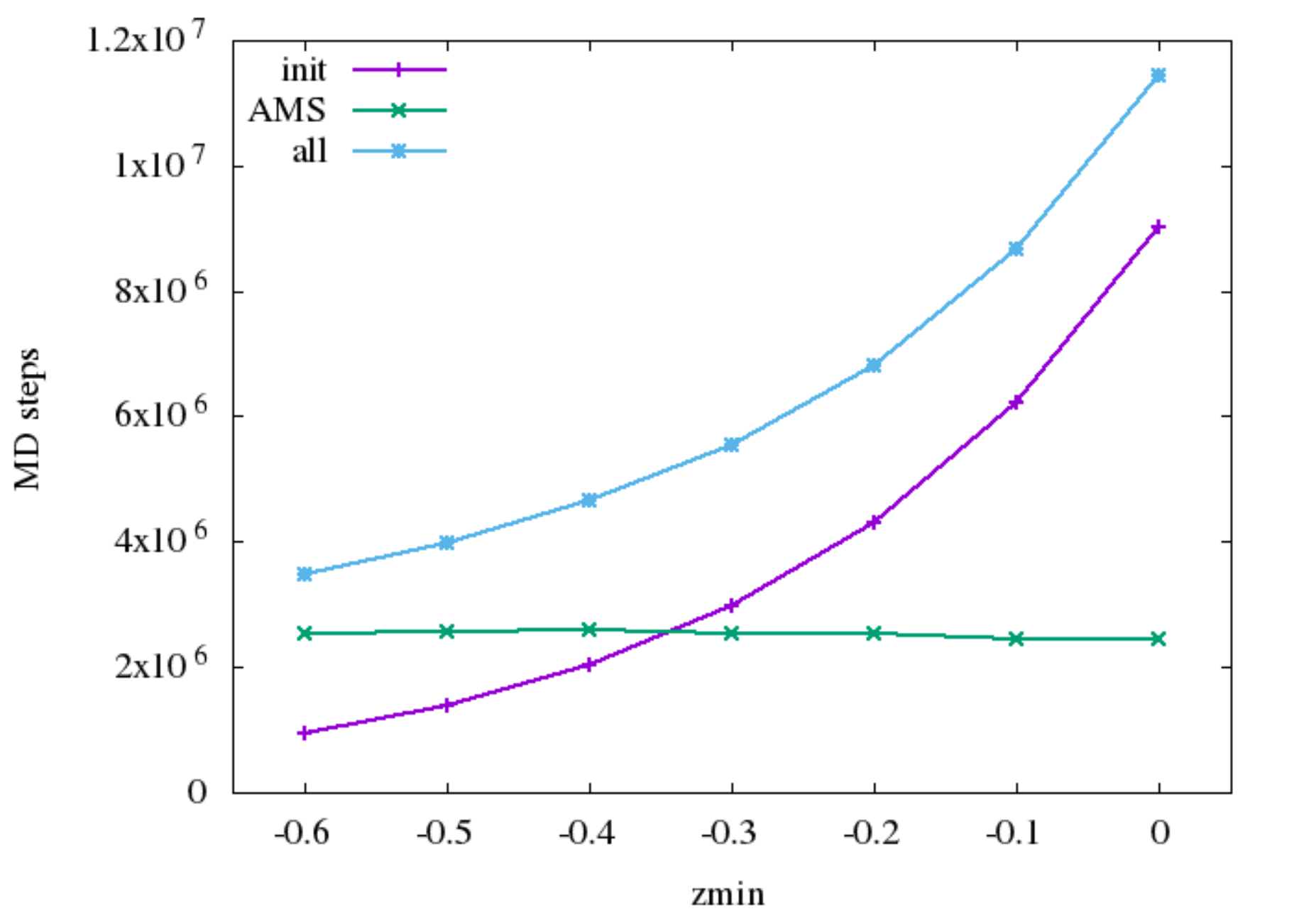}
\caption{
Simulation steps used to initiate the 500 replicas and for each AMS run.
}
\label{cond_var_steps}
\end{figure}

Another important feature to be considered when fixing $z_{min}$ is the time required to initiate the replicas and to run the AMS simulations.
This is shown in Figure \ref{cond_var_steps}.
The time for the initiation phase tends to grow exponentially as $z_{min}$ is larger.
However, because the AMS method is appropriate to simulate rare events, the AMS simulation time is approximately constant.
Thus, we conclude it is better to have $\Sigma_{z_{min}}$ closer to $A$.

\subsection{Calculating the committor function}
\label{section:committor}

Another quantity of interest is the committor function:
\begin{equation}
p(x)=\mathbb{P}(\tau_B<\tau_A | X_0=x),
\end{equation}
i.e. the probability of entering $A$ before $B$ when starting from $x$. 
Note that, from the definition of a conditional probability, it is possible to rewrite $p(x)$ as:
\begin{equation}
p(x)=\frac{p_{B,X_0}(x)}{p_{X_0}(x)}=\frac{\mathbb{P}(\tau_B<\tau_A \cap X_0=x)}{\mathbb{P}(X_0=x)}.
\label{eq:defcommittor}
\end{equation}

To approximate the committor function let us consider a large set of $N$ trajectories $(\mathbf{X}_{t\in [0,\tau_{AB}^n]})_{1\leq n \leq N}$ at equilibrium that starts outside $A$ and $B$.
Using the same strategy as for the flux, the space is split into $L$ cells $(C_l)_{1\leq l \leq L}$.
Let us now introduce an approximation of the numerator $p_{B,X_0}(x)$ and the denominator $p_{X_0}(x)$ in Equation~\eqref{eq:defcommittor}, for each cell $C_l$:
\begin{equation}
p_{B,X_0}(C_l)=\frac{\displaystyle \sum_{n=1}^N \mathds{1}_{\tau_B^n<\tau_A^n} \sum_{t=0}^{\tau_{AB}^n} \mathds{1}_{X_t^n \in C_l}}{\displaystyle \sum_{n=1}^N (\tau_{AB}^n+1)},
\label{eq:pBX}
\end{equation}
\begin{equation}
p_{X_0}(C_l)=\frac{\displaystyle \sum_{n=1}^N \sum_{t=0}^{\tau_{AB}^n} \mathds{1}_{X_t^n \in C_l}}{\displaystyle \sum_{n=1}^N (\tau_{AB}^n+1)}.
\label{eq:pX}
\end{equation}
Note that this consists in counting each time a trajectory passes through $C_l$ for $p_{X_0}(C_l)$ and considering it in $p_{B,X_0}(C_l)$ only if the trajectory enters $B$ before $A$.
Since we consider trajectories at equilibrium, $p_{B,X_0}(C_l)$ (resp. $p_{X_0}(C_l)$) actually approximates the probability to reach $B$ before $A$ and to be in $C_l$ (resp. the probability to be in $C_l$) for a trajectory starting at equilibrium in $C_l$.

Let us now consider $M$ AMS runs, where a total of $N_m$ replicas $\mathbf{X}_{t\in [0,\tau_{AB}^{n,m}]}^{n,m}$ where obtained for each run $m$, and call $w_{n,m}$ the weight of $n^{\text th}$ replica from the $m^{\text th}$ run.
From Equation~\eqref{eq:any_ams}, the following approximations for Equations~\eqref{eq:pBX}~and~\eqref{eq:pX} are obtained:
\begin{equation}
\tilde{p}_{B,X_0}(C_l)= \frac{\displaystyle \sum_{m=1}^M \sum_{n=1}^{N_m} w_{n,m} \mathds{1}_{\tau_B^{n,m}<\tau_A^{n,m}} \sum_{t=0}^{\tau_{AB}^{n,m}} \mathds{1}_{X_t^{n,m} \in C_l}}{ \displaystyle \sum_{m=1}^M \sum_{n=1}^{N_m} w_{n,m} (\tau_{AB}^{n,m}+1)}
\label{eq:tpBX0}
\end{equation}
\begin{equation}
\tilde{p}_{X_0}(C_l)=\frac{\displaystyle \sum_{m=1}^M \sum_{n=1}^{N_m} w_{n,m} \sum_{t=0}^{\tau_{AB}^{n,m}} \mathds{1}_{X_t^{n,m} \in C_l}}{\displaystyle \sum_{m=1}^M \sum_{n=1}^{N_m} w_{n,m} (\tau_{AB}^{n,m}+1)}
\label{eq:tpX0}
\end{equation}
The division of \eqref{eq:tpBX0} by \eqref{eq:tpX0} gives us an estimation $\tilde{p}(C_l)$ of the committor function in cell $C_l$:
\begin{equation}
\tilde{p}(C_l)=\frac{\displaystyle \sum_{m=1}^M \sum_{n=1}^{N_m} w_{n,m} \mathds{1}_{\tau_B^{n,m}<\tau_A^{n,m}} \sum_{t=0}^{\tau_{AB}^{n,m}} \mathds{1}_{X_t^{n,m} \in C_l}}{\displaystyle \sum_{m=1}^M \sum_{n=1}^{N_m} w_{n,m} \sum_{t=0}^{\tau_{AB}^{n,m}} \mathds{1}_{X_t^{n,m} \in C_l}}.
\label{eq:committor-final}
\end{equation}

\begin{figure}[htb!]
\centering
\includegraphics[width=0.9\linewidth,trim={2cm 4cm 1cm 6.2cm},clip]{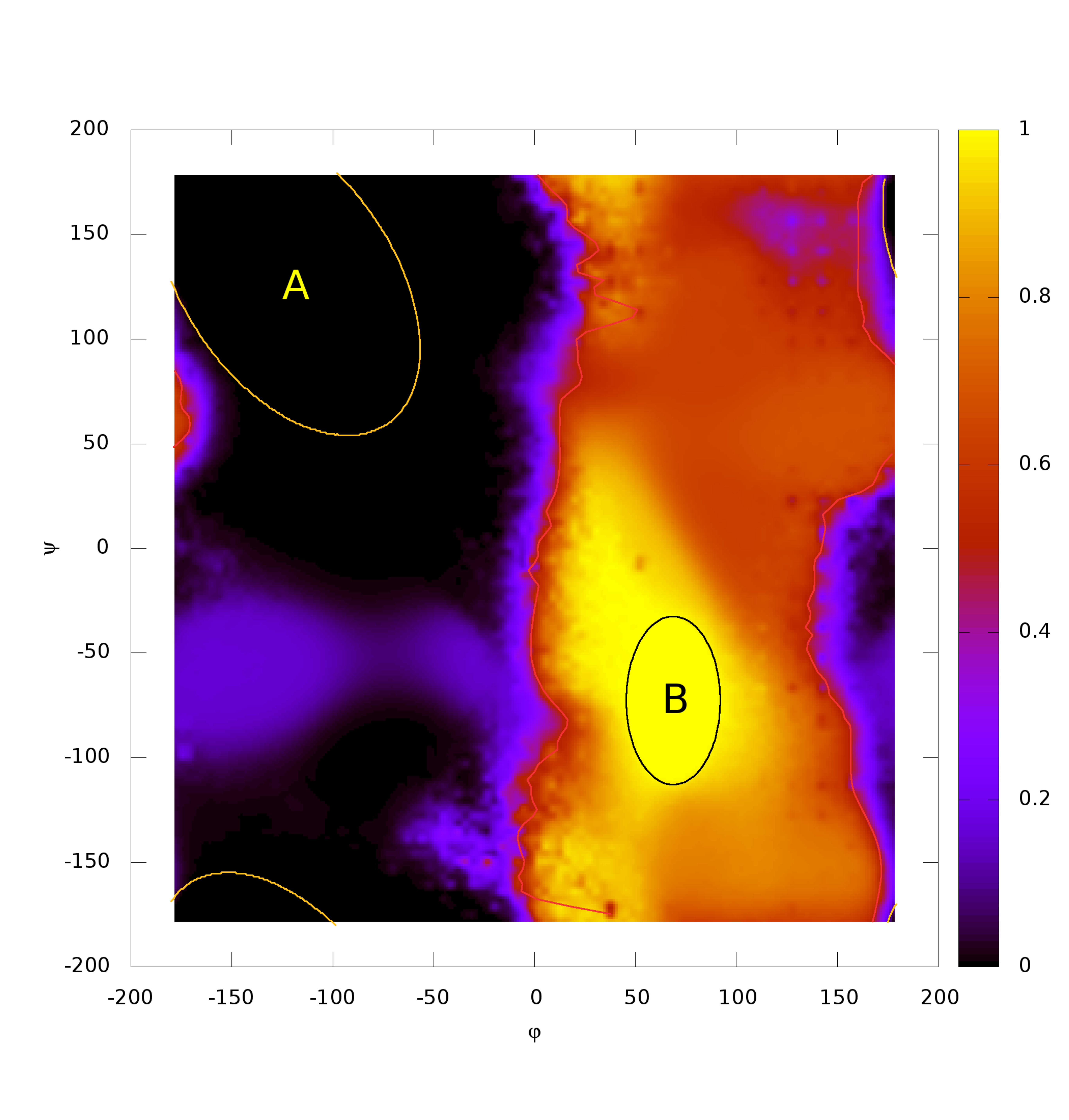}
\includegraphics[width=0.9\linewidth,trim={2cm 4cm 1cm 6.2cm},clip]{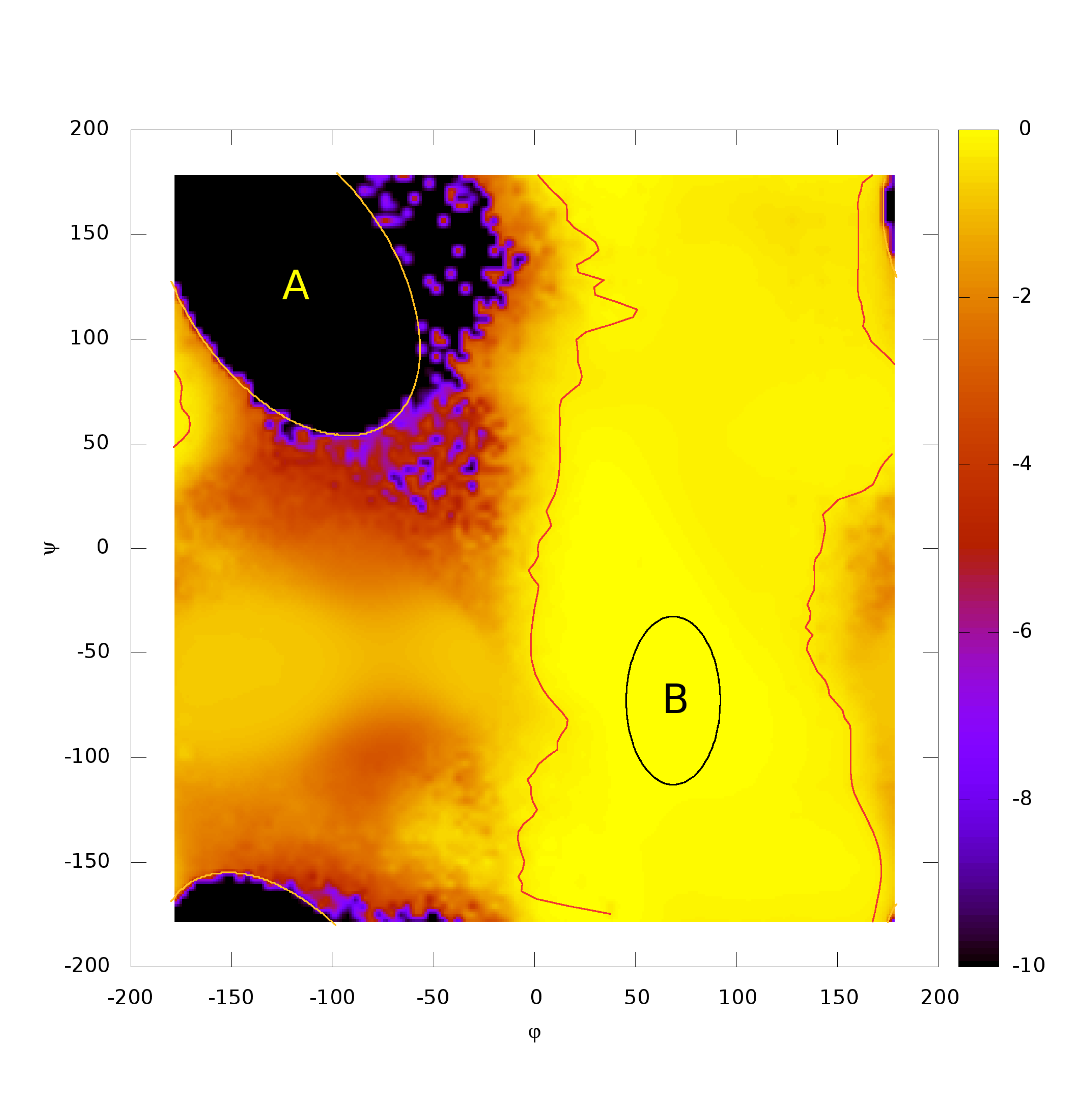}
\caption{
The committor function obtained with 5000 AMS runs with 100 replicas each.
In the second figure the same result is presented in log-scale, with a cut at $10^{-10}$.
We used initial conditions at equilibrium, starting from equally distributed $(\varphi,\psi)$ positions over the Ramachandran plot.
The red lines mark the isolevel $0.5$, where the probability to enter $A$ before $B$ is the same as to enter $B$ before $A$, namely the transition state.
}
\label{committor}
\end{figure}

The result obtained using Equation~\eqref{eq:committor-final} is given in Figure \ref{committor}.
\FloatBarrier

%

\section*{Acknowledgments}
The authors would like to thank Najah-Imane Bentabet who worked on a preliminary version of the AMS algorithm for the NAMD code, and J\'er\^ome H\'enin for fruitful discussions.
Part of this work was completed while the authors were visiting IPAM during the program "Complex High-Dimensional Energy Landscapes".
The authors would like to thank IPAM for its hospitality.
This work is supported by the European Research Council under the European Union's Seventh Framework Programme (FP/2007-2013)/ERC Grant Agreement number 614492.

\end{document}